\documentclass[preprint]{elsarticle}

\usepackage{epsf}
\usepackage{epsfig}
\usepackage{amsmath}
\usepackage{dsfont}
\usepackage{graphicx}
\usepackage{booktabs}
\usepackage{float}
\usepackage{slashed}

\newcommand{\Op}{\mathcal{O}}  
\newcommand{\eins}{\mathds{1}} 

\usepackage{color}

\def\beq{\begin{equation}}
\def\eeq{\end{equation}}
\def\bea{\begin{eqnarray}}
\def\eea{\end{eqnarray}}
\def\beqa{\begin{equation}\begin{array}{l}}
\def\eeqa{\end{array}\end{equation}}
\def\eqlab#1{\label{eq:#1}}
\def\figlab#1{\label{fig:#1}}
\def\tablab#1{\label{tab:#1}}

\def\eref#1{(\ref{eq:#1})}
\def\Eqref#1{Eq.~(\ref{eq:#1})}

\def\Figref#1{Fig.~\ref{fig:#1}}

\def\half{\mbox{\small{$\frac{1}{2}$}}}

\def\al{\alpha}
\def\be{\beta}
\def\ga{\gamma} 
 
\def\veps{\varepsilon}  \def\eps{\epsilon}

\def\la{\lambda}

\def\pa{\partial}
\def\vrho{\varrho}

\def\pa{\partial}

\def\nn{\nonumber}

\def\mathscr{\mathcal}

\journal{Nuclear Physics B}

\begin{document}

\title{Quark transverse charge densities in the $\Delta(1232)$ from lattice QCD}

\author[nic]{Constantia Alexandrou}
\author[nic]{Tomasz Korzec}
\author[nic]{Giannis Koutsou}
\author[mai]{C\'edric Lorc\'e}
\author[mit]{John W. Negele}
\author[mai]{Vladimir Pascalutsa}
\author[ath]{Antonios Tsapalis}
\author[mai]{Marc Vanderhaeghen}

\address[nic]{Department of Physics, University of Cyprus, P.O. Box 20537, 1678
Nicosia, Cyprus}
\address[mai]{Institut f\"ur Kernphysik, Johannes Gutenberg-Universit\"at,
D-55099 Mainz, Germany}
\address[mit]{Center for Theoretical Physics, Laboratory for Nuclear Science
  and Department of Physics, Massachusetts Institute of Technology, Cambridge,
Massachusetts 02139, U.S.A.}
\address[ath]{Institute of Accelerating Systems and Applications, University of
Athens, Athens, Greece}

\date{\today}

\begin{abstract}

We extend the formalism relating electromagnetic form factors to  transverse  quark charge densities 
in the light-front frame to 
the case of a spin-3/2 baryon and calculate these transverse densities for  the $\Delta(1232)$ isobar using lattice QCD.
The transverse charge densities for a transversely polarized spin-3/2 particle are characterized by 
monopole, dipole, quadrupole, and octupole patterns representing the structure beyond that of a pure point-like spin-3/2 particle. 
We present lattice QCD results for the $\Delta$-isobar electromagnetic 
form factors for pion masses down to approximatively 350 MeV for three 
cases: quenched QCD, two-degenerate flavors of dynamical Wilson quarks, 
and three 
flavors of quarks using a mixed action that combines domain wall valence 
quarks and dynamical staggered sea quarks. 
We extract transverse quark charge densities from these lattice results and find that the $\Delta$ is  prolately deformed, as indicated by the fact that the quadrupole moment $G_{E2}(0$) is larger than the value $-3$ characterizing a point particle and the fact that the transverse charge density in a 
$\Delta^+$ of maximal transverse spin projection  
is elongated along the axis of the spin. 

\end{abstract}

\begin{keyword} 
Electromagnetic form factors, Electric and Magnetic Moments, $\Delta$-resonance  
\PACS 13.40.Gp \sep 13.40.Em \sep 14.20.Gk
\end{keyword}

\maketitle

\section{Introduction}

The question of how the structure of mesons and baryons arises from the 
interaction among their  quark and gluon constituents is at the forefront of 
contemporary research in hadron physics. Key observables in this field 
are the electromagnetic (e.m.) form factors (FFs), which yield
the distribution of the quark charges in a hadron. The nucleon
e.m.\ FFs have been especially thoroughly studied recently at  electron-beam
facilities, such as Jefferson Lab, MIT-Bates, and MAMI, 
see Refs.~\cite{HydeWright:2004gh,Arrington:2006zm,Perdrisat:2006hj} for 
latest reviews. 

In systems like nuclei or atoms, in which the spatial size $R$ is large compared to the system's Compton wavelength $1/M$, electromagnetic form factors are  essentially three dimensional  Fourier transforms (or distorted wave analogs thereof) of ground state charge density distributions, and thereby provide valuable physical insight into the structure of the ground state of the system in the lab frame.  In hadrons, however,  where $ R \sim 1/M $, the Fourier transform argument is inapplicable, and there is no known way to relate form factors to charge densities in the lab frame. 
However, when the hadron  is viewed from a light front, there is a simple and consistent field-theoretic density interpretation of FFs 
as the Fourier transform of  the spatial distribution of the quark charge 
in the plane transverse to the line-of-sight.  Equivalently, in the infinite momentum frame, the quantity that plays the role of $M$ in the lab frame Fourier transform argument is 
the light-front momentum $p^+$, which goes to infinity, so the
form factor is just the two-dimensional Fourier transform of the transverse density in the infinite momentum frame~\cite{Burkardt:2000za}. 
In this way, the transverse quark charge densities have been mapped out
in the nucleon~\cite{Miller:2007uy,Carlson:2007xd}, 
 deuteron~\cite{Carlson:2008zc}, and pion~\cite{Miller:2009qu} based on empirical FFs. 
 It is important to note, however, that physics in the lab and infinite momentum frames is significantly different and hence there is no simple relation between the two-dimensional transverse densities in the infinite momentum (or light front) frame and the three-dimensional density in the lab frame, and we will return to this issue in the discussion of deformation. 

In this work we address the e.m.\ FFs and transverse charge densities
of the $\Delta(1232)$ isobar, the lightest nucleon excitation. 
The nucleon-to-$\Delta$ transition
has been well measured experimentally over a large range of photon
virtualities, see Ref.~\cite{Pascalutsa:2006up} for a recent review. 
It is dominated by a magnetic dipole 
transition, while the electric and Coulomb quadrupole transitions
were found to be small (in the few percent range as compared to
the magnetic dipole transition). These measurements have made it possible  
to quantify the deformation of the 
$N \to \Delta$ transition charge distribution~\cite{Carlson:2007xd}. 

On the other hand, the information on the e.m.\ FFs of the 
$\Delta$ itself is scarce. Because of the tiny lifetime
of the $\Delta$, it is of course very difficult if not impossible 
to access these quantities directly in experiment.
Fortunately, however, as in the case of the $N \to \Delta$ FFs, it is now feasible to 
calculate the $\Delta$ FFs using lattice QCD. 
First lattice QCD results in the quenched approximation  were presented 
in Ref.~\cite{Alexandrou:2007dt}, and first results on the $\Delta$ FFs 
using dynamical quarks were
 briefly reported 
in~\cite{Alexandrou:2008bn}.
In this more extended publication we provide further details 
on the lattice evaluation of these FFs as well as
the interpretation of the calculated FFs in 
terms of the transverse charge densities. 

The outline of the paper is as follows:
Section~\ref{sec2}, contains a general discussion of the e.m.\ interaction of a 
spin-3/2 system along with the definition of its e.m.\ moments and FFs. 
In Section~\ref{sec3}, we find the specific (`natural') values for electromagnetic moments
that an elementary (pointlike) spin-3/2 particle would possess. 
In Section~\ref{sec4}, we introduce the light-front helicity amplitudes 
for the e.m. vertex of a spin-3/2 system, and express them in terms of the 
corresponding e.m. FFs. In Section~\ref{sec5}, we calculate quark transverse 
charge densities of a spin-3/2 system in terms of the light-front helicity 
amplitudes. We also calculate the values of the electric dipole, quadrupole and
octupole moments of these transverse charge densities, and show that for a 
pointlike spin-3/2 particle they vanish. 
In Section~\ref{sec6}, we describe our three different types of lattice calculations  of the 
$\Delta(1232)$ e.m. FFs and compare their results. 
In the first type,  we use Wilson 
fermions in the quenched approximation. The second type uses 
two-degenerate flavors  of dynamical Wilson quarks ($N_F=2$). 
The third type uses a 
mixed action combining domain wall valence quarks with  
dynamical staggered sea quarks including light degenerate up and down quarks and heavier strange quarks ($N_F=2+1$). 
In Section~\ref{sec7}, we use these lattice QCD results to 
extract the transverse  quark charge densities in the $\Delta$ resonance. 
We summarize our results, discuss their relation to models and present our conclusions in 
 Section~\ref{sec8}. The lattice results for the $\Delta$ 
e.m. FFs are tabulated in an Appendix.

\section{The $\gamma^* \Delta \Delta$ vertex and form factors}
\label{sec2}

\begin{figure}[h]
\centerline{
\epsfxsize=7cm
\epsffile{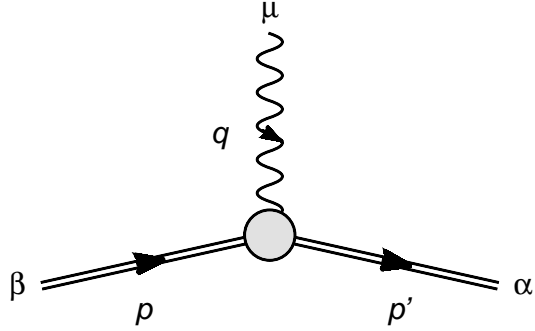}
}
\caption{The $\gamma^* \Delta \Delta$ vertex. The four-momenta of 
the initial (final) $\Delta$ and of the photon are given by 
$p$ ($p^\prime$) and $q$ respectively. 
The four-vector indices of the initial (final) 
spin-3/2 fields are given by $\beta$ ($\alpha$), and 
$\mu$ is the four-vector index of the photon field.}
\label{fig:gadeldeltreevertex}
\end{figure}

Let us consider the coupling of a photon to a $\Delta$,
shown in Fig.~\ref{fig:gadeldeltreevertex}.  
The matrix element of the electromagnetic current operator $J^\mu$ 
between spin-3/2 states can be decomposed into four multipole transitions:
a Coulomb monopole (E0), a magnetic dipole (M1), 
a Coulomb quadrupole (E2) and 
a magnetic octupole (M3).
We first write a Lorentz-covariant decomposition for the 
on-shell $\gamma^* \Delta \Delta$ vertex,  which exhibits
manifest electromagnetic 
gauge-invariance~\cite{Pascalutsa:2006up,Nozawa:1990gt}: 
\begin{eqnarray}
&&\langle \Delta(p',\lambda^\prime) \,|\,J^\mu(0) \,|\, \Delta (p,\lambda) 
\rangle  \nonumber \\
&&= -  \bar u_\alpha (p',\lambda^\prime) \left\{  \left[  
F_1^\ast(Q^2)  g^{\alpha \beta}  
+ F_3^\ast(Q^2) \frac{q^\alpha q^\beta}{(2 M_\Delta)^2} 
\right] \gamma^\mu \right. \nonumber \\
&& +\left. \left[ F_2^\ast(Q^2)  g^{\alpha \beta}
+ F_4^\ast(Q^2) \frac{q^\alpha q^\beta}{(2M_\Delta)^2}\right] 
\frac{i \sigma^{\mu\nu} q_\nu}{2 M_\Delta} \, \right\} u_\beta(p,\lambda) ,
\label{eq:gadeldeltree}
\end{eqnarray}
where $M_\Delta = 1.232$~GeV is the $\Delta$ mass, 
$u_\alpha$ is the Rarita-Schwinger spinor for a spin-3/2 state, 
and $\lambda$ ($\lambda^\prime$) are the initial (final) $\Delta$ helicities.  
Furthermore, $F^\ast_{1,2,3,4}$ 
are the $\gamma^* \Delta \Delta$ form factors, and 
$F^\ast_1(0) = e_\Delta$ is the $\Delta$ electric charge in units of $e$ 
(e.g., $e_{\Delta^+} = +1$). 
For  future reference, we also define the quantity 
$\tau \equiv Q^2 / (4 M_\Delta^2)$. 

A physical interpretation of the four electromagnetic $\Delta \to \Delta$ 
transitions can be obtained by performing a multipole 
decomposition~\cite{Weber:1978dh,Nozawa:1990gt}. 
For this purpose it is convenient to consider the Breit frame, where 
$\vec p$ = $- \vec p^{\, \prime}$ = $- \vec q / 2$. Furthermore, we choose 
$\vec q$ along the $z$-axis and denote the initial (final) $\Delta$ spin
projections along the $z$-axis by $s$ ($s^\prime$).  
In this frame, the matrix elements of the charge operator define  
the Coulomb monopole (charge) and Coulomb quadrupole form factors as~:
\begin{eqnarray}
\langle \frac{\vec q}{2} , s^\prime \, |\, J^0(0) \,|\, 
- \frac{\vec q}{2}, s \rangle &\equiv& 
(2 M_\Delta) \, \delta_{s^\prime \, s} \, 
\left\{ 
\left( \delta_{s\, \pm \frac{3}{2} } + \delta_{s\, \pm \frac{1}{2} }
\right) \, G_{E0} (Q^2)  \right. 
\nonumber \\
&&\left. \hspace{2cm} -  \frac{2}{3} \tau   
\left( \delta_{s\, \pm \frac{3}{2} } - \delta_{s\, \pm \frac{1}{2} }
\right) \, G_{E2} (Q^2)
\right\}.  
\label{eq:breit0}
\end{eqnarray}  
Using Eq.~(\ref{eq:gadeldeltree}), we can express the 
Coulomb monopole and quadrupole 
form factors in terms of $F^\ast_{1,2,3,4}$ as~:
\begin{eqnarray}
G_{E0} &=& \left( F_1^\ast - \tau F_2^\ast \right) 
+ \frac{2}{3} \tau G_{E2}, \\ 
\label{eq:e0}
G_{E2} &=& \left( F_1^\ast - \tau F_2^\ast \right) 
- \frac{1}{2} ( 1 + \tau) 
\left( F_3^\ast - \tau F_4^\ast \right).  
\label{eq:e2}
\end{eqnarray}
Analogously,  the matrix elements of the current operator 
define the magnetic dipole and magnetic octupole form factors. 
For the transverse spherical component 
$J_{+1} \equiv - \frac{1}{\sqrt{2}} ( J^1 + i J^2)$ we obtain~:
\begin{eqnarray}
&&\langle \frac{\vec q}{2} , s^\prime \, |\, J_{+1} (0) \,|\, 
- \frac{\vec q}{2}, s \rangle 
\equiv 
(- \sqrt{2}) (2 M_\Delta) \, \frac{\sqrt{\tau}}{\sqrt{3}} \, 
\nonumber \\
&& \hspace {1cm} \times \left\{ 
\left( \delta_{s^\prime \, +\frac{3}{2} } \, \delta_{s\, +\frac{1}{2} }
+ \delta_{s^\prime \, -\frac{1}{2} } \, \delta_{s\, -\frac{3}{2} }
+ \frac{2}{\sqrt{3}} 
\delta_{s^\prime\, +\frac{1}{2} } \, \delta_{s\, -\frac{1}{2} }
\right) \, G_{M1} (Q^2) \right. \nonumber \\
&&\left. \hspace{1cm} -  \frac{4}{5} \tau   
\left( \delta_{s^\prime \, +\frac{3}{2} } \, \delta_{s\, +\frac{1}{2} }
+ \delta_{s^\prime \, -\frac{1}{2} } \, \delta_{s\, -\frac{3}{2} }
- \sqrt{3} 
\delta_{s^\prime\, +\frac{1}{2} } \, \delta_{s\, -\frac{1}{2} }
\right) \, G_{M3} (Q^2) 
\right\}.  
\label{eq:breiti}
\end{eqnarray}  
Using Eq.~(\ref{eq:gadeldeltree}), we can express the 
magnetic dipole and octupole form factors in terms of $F^\ast_{1,2,3,4}$ as~:
\begin{eqnarray}
G_{M1} &=& \left( F_1^\ast + F_2^\ast \right) 
+ \frac{4}{5} \tau G_{M3}, \\
\label{eq:m1}
G_{M3} &=& \left( F_1^\ast + F_2^\ast \right) 
- \frac{1}{2} (1 + \tau) \left( F_3^\ast + F_4^\ast \right). 
\label{eq:m3}
\end{eqnarray}
At $Q^2 = 0$, the multipole form factors define 
the charge ($e_\Delta$), the magnetic dipole moment ($\mu_\Delta$),
the electric quadrupole moment ($Q_\Delta$),  
and the magnetic octupole moment ($O_\Delta$) as~:
\begin{subequations}
\begin{eqnarray}
\eqlab{EMmoments}
e_\Delta &=& G_{E0}(0) = F^\ast_1(0), \\
\mu_\Delta &=& \frac{e}{2 M_\Delta} G_{M1}(0) = 
\frac{e}{2 M_\Delta} \left[e_\Delta+F_2^\ast (0)\right]  \,,\\
Q_\Delta & = &  \frac{ e}{ M_\Delta^2} G_{E2}(0) = 
\frac{ e}{ M_\Delta^2}\left[ e_\Delta - 
\frac{1}{2} F_3^\ast(0)\right]\,, \label{eq:ee8c}\\
O_\Delta & = &  \frac{e}{2M_\Delta^3} G_{M3}(0) = 
\frac{e}{2M_\Delta^3}
\left[ e_\Delta + F_2^\ast (0) - \frac{1}{2} 
\left(F_3^\ast(0)+F_4^\ast(0)\right)\right]\,.
\label{eq:ee8}
\end{eqnarray}
\end{subequations}
In the following, we will also use the relations that
express the form factors $F^\ast_{1,2,3,4}$ 
in terms of the multipole form factors~:
\begin{eqnarray}
F^\ast_1&=&\frac{1}{1 + \tau} 
\left\{ G_{E0} - \frac{2}{3} \tau G_{E2}  
+ \tau \Big[ G_{M1} - \frac{4}{5} \tau G_{M3} \Big] \right\}, 
\nonumber \\
F^\ast_2&=&-\frac{1}{1 + \tau} 
\left\{ G_{E0} - \frac{2}{3} \tau G_{E2}  
- \Big[ G_{M1} - \frac{4}{5} \tau G_{M3} \Big] \right\}, 
\nonumber \\
F^\ast_3&=&\frac{2}{(1 + \tau)^2} 
\left\{ G_{E0} - \left( 1 + \frac{2}{3} \tau \right)  G_{E2}  
+ \tau \Big[ G_{M1} - \left( 1 + \frac{4}{5} \tau \right) G_{M3} \Big] 
\right\}, \nonumber \\
F^\ast_4&=&\!- \frac{2}{(1 + \tau)^2} 
\left\{ G_{E0} - \!\left( 1 + \frac{2}{3} \tau \right)  G_{E2}  
- \Big[ G_{M1} -\! \left( 1 + \frac{4}{5} \tau \right) G_{M3} \Big] 
\right\}.
\end{eqnarray}
For completeness we also express the form factors  in terms of the
covariant vertex functions $a_1,a_2,c_1$ and $c_2$, used for
instance in references~\cite{Nozawa:1990gt,Leinweber:1992hy,Alexandrou:2008bn}~:
\begin{equation}
   F_1^\ast = a_1 + a_2\, , \qquad F_2^\ast = -a_2\, , \qquad
   F_3^\ast = c_1 + c_2\, , \qquad F_4^\ast = -c_2\, .
\end{equation}
These relations, together with the identity
\begin{eqnarray}
\bar u_\alpha(p',\lambda')i\sigma^{\mu\nu}q_\nu u_\beta(p,\lambda) = \bar
u_\alpha(p',\lambda')\left( 2M_\Delta\gamma^\mu - [p+p']^\mu \right) 
u_\beta(p,\lambda)
\end{eqnarray}
bring Eq.~(\ref{eq:gadeldeltree}) into the form introduced 
in~\cite{Nozawa:1990gt,Leinweber:1992hy,Alexandrou:2008bn}.

The empirical knowledge of the $\Delta$ electromagnetic moments  
is scarce, even though there were several attempts to measure the 
magnetic moment.  
The current Particle Data Group value of the $\Delta^+$ 
magnetic dipole moment is given as~\cite{PDG2008}:
\beq
\mu_{\Delta^+} =  2.7 \mbox{${{+1.0} \atop {-1.3}}$}
(\mathrm{stat.}) \pm 1.5 (\mathrm{syst.}) \pm 3 (\mathrm{theor.})\, \mu_N\,,
\label{eq:mdmex}
\eeq
where  $\mu _N=e/2M_N$ is the nuclear magneton. This result was  obtained from 
{\it radiative photoproduction}~($\gamma N \to \pi N \gamma^\prime$) of
neutral pions in the $\Delta(1232)$ region by the TAPS Collaboration at
MAMI~\cite{Kotulla:2002cg}, using 
a phenomenological model of the $\gamma p \to \pi^0 p \gamma^\prime$ 
reaction~\cite{Drechsel:2001qu}. 
For the $\Delta^+$, Eq.~(\ref{eq:mdmex}) implies~: 
\beq
G_{M1}(0) =  3.5 \mbox{${{+1.3} \atop {-1.7}}$}
(\mathrm{stat.}) \pm 2.0 (\mathrm{syst.}) \pm 3.9 (\mathrm{theor.})\, .
\eeq
The size of the error-bar is
rather large due to both experimental and theoretical uncertainties.
Recently a dedicated experimental effort
is underway at MAMI using the Crystal Ball detector~\cite{Kotulla:2007zz}, 
aiming at
improving on the statistics of the TAPS data by almost
two orders of magnitude.

For the $\Delta$ electric quadrupole moment, no direct measurements exist. 
However the electric quadrupole moment for the $N \to \Delta$ 
transition has been measured accurately from the 
$\gamma N \to \pi N$ reaction at the $\Delta$ resonance 
energy~\cite{Tiator:2003xr}~:
\begin{eqnarray}
Q_{p \to \Delta^+} \,&=&\, - \left( 0.0846 \pm 0.0033 \right ) 
\; \mathrm{e} \cdot \mathrm{fm}^2.
\label{eq:qndelexp}
\end{eqnarray} 
In the large-$N_c$ limit of QCD, the two are related 
as follows~\cite{Buchmann:2002mm}:
\begin{eqnarray}
\frac{Q_{\Delta^+}}{Q_{p \to \Delta^+}} \,=\, \frac{2 \sqrt{2}}{5} \,+\, 
{\mathcal O} \left(\frac{1}{N_c^2} \right), 
\label{eq:qlargenc}
\end{eqnarray}
which, using the empirical value for $Q_{p \to \Delta^+}$ 
yields for the $\Delta^+$ quadrupole moment:
\begin{eqnarray}
Q_{\Delta^+} \,&=&\, - \left( 0.048 \pm 0.002 \right ) \; 
\mathrm{e} \cdot \mathrm{fm}^2,
\label{eq:qdellargenc}
\end{eqnarray} 
accurate up to corrections of order $1/N_c^2$. 
Note that using Eq.~(\ref{eq:ee8}), the value in Eq.~(\ref{eq:qdellargenc}) implies
$G_{E2}(0) = -1.87 \pm 0.08$.

\section{The `natural' values of the e.m. moments}
\label{sec3}

As first argued by Weinberg~\cite{Weinberg:1970bu}, based on the 
Gerasimov-Drell-Hearn sum rule,
there is a `natural' value for
the magnetic moment of a pointlike particle with spin, which 
corresponds to a gyromagnetic ratio $g$ equal to 2.   
 It has later been observed
that all consistent field theories of charged particles 
with spin respect this value, see e.g.~\cite{Ferrara:1992yc,Holstein:2006wi}. 
Given this universal result for the magnetic moment, it is reasonable to expect 
that all electromagnetic moments of pointlike particles are fixed at `natural' values. 
To determine the values for the spin-3/2 case, we examine the values of
the e.m. moments of the gravitino in 
extended supergravity~\cite{Ferrara:1976fu,VanNieuwenhuizen:1981ae}.

The gravitino, if it existed,  would be a spin-3/2 particle described 
by a Rarita-Schwinger field 
which, in the framework  of ${\mathcal N}=2$ supergravity, couples
consistently to electromagnetism. 
We therefore expect that all the e.m.\ moments arising in this
theory are `natural'. 
To find their values, we start from the
following Lagrangian density:\footnote[1]{In our conventions: 
$\ga^{\mu\nu} = \half [\ga^\mu, \ga^\nu]$,
 $\ga^{\mu\nu\al} = \half \{ \ga^{\mu\nu}, \ga^\al\}$, 
 $F^{\mu\nu}=\pa^{\mu} A^{\nu}- \pa^{\nu} A^{\mu}$, $\tilde  F^{\mu\nu} =
\veps^{\mu\nu\vrho\la} \pa_\vrho A_\la$.}
\bea
{\mathcal L} & = &  \bar \psi_{\mu}\, \ga^{\mu\nu\al} (i\pa_\al-e A_\al)\,\psi_\nu - m \,\bar \psi_{\mu} \ga^{\mu\nu}
\psi_\nu \nn\\
& + & em^{-1}\,
\bar \psi_{\mu}\, ( \, i \kappa_1 F^{\mu\nu} - \kappa_2  \ga_5 \tilde F^{\mu\nu}  \,) \, \psi_\nu\,.
\eqlab{emRS}
\eea
It describes the spin-3/2 Rarita-Schwinger field ($\psi_\mu$)
with mass $m$ coupled to the electromagnetic field ($A_\mu$) via the minimal coupling
(with positive charge $e$) and two non-minimal couplings $\kappa_1$ and $\kappa_2$. 
As shown in Ref.~\cite{Deser:2000dz}, this is the most general Lagrangian
of this type
that gives the right number of spin degrees of freedom for a spin-3/2 particle. 
This theory, however, would still lead to rather subtle pathologies 
such as non-causal wave
propagation~\cite{Johnson:1961vt, Velo:1969bt}, at least if no other fields are present. Adding gravity
in a supersymmetric way makes the theory fully consistent from this viewpoint, but 
also constrains the non-minimal couplings as follows:
\beq
 \kappa_1 = \kappa_2 = 1.
 \eqlab{sugrachoice}
\eeq
We shall refer to these values as the `SUGRA choice'.

The e.m.\ vertex stemming from~\Eqref{emRS} is 
\beq
\Gamma^{\al\be\mu} (p', p) = \ga^{\al\be\mu} - \frac{\kappa_1}{m}
(q^\al g^{\be\mu} -   q^\be g^{\al\mu}) + i  
\frac{\kappa_2}{m} \gamma_5 \veps^{\al\be\mu\vrho} q_\vrho\,,
\eeq
where $q=p'-p$, and $\varepsilon_{0123} = +1$. 
It is easy to verify that this coupling conserves the e.m. current:
\beq
q_\mu \, \bar u_\al (p')\, \Gamma^{\al\be\mu} (p', p)\, u_\be(p) = 0\,,
\eeq
as well as, for the SUGRA choice, the supersymmetric current:
\begin{subequations}
\bea
(p_\al' - \half m \ga_\al)  \, \Gamma^{\al\be\mu} (p', p)\, u_\be(p) \, \eps_\mu (q) & = & 0\, ,\\
   \bar u_\al (p')\, \Gamma^{\al\be\mu} (p', p)\, (p_\be - \half m \ga_\be) \,
   \eps_\mu (q) & = & 0\,,
\eea
\label{eq:super}
\end{subequations}
where  $\eps_\mu (q)$ is the photon polarization vector, and $q\cdot \eps = 0
= q^2$ in Eqs.~(\ref{eq:super}).

The matrix elements of this vertex can be compared with the general decomposition
of the spin-3/2 e.m.\  current given in \Eqref{gadeldeltree}, and in doing so we obtain
the following result:
\begin{subequations}
\bea
F_1^\ast & = & 1 + 2 (\kappa_1 +\kappa_2) \,\tau\stackrel{\mathrm{\sl sugra}}{=} 1+4\tau\,, \\
F_2^\ast &=& 2 \kappa_1\stackrel{\mathrm{\sl sugra}}{=} 2\,,\\
F_3^\ast & = & 4  (\kappa_1 +\kappa_2) \stackrel{\mathrm{\sl sugra}}{=} 8\,,\\
F_4^\ast & = & 0\,,
\eea
\label{eq:sugravalues}
\end{subequations}
Thus, the values of gravitino's
e.m.\ moments in 
$\mathcal{N}=2$ supergravity are: 
\beq
G_{E0} (0)= 1, \,\,\, G_{M1} (0)= 3, \,\,\, G_{E2}(0) = -3, \,\,\,G_{M3} (0)=
-1.
\label{eq:natvalues}
\eeq 
We take these values as the  `natural' values characterizing a structureless spin-3/2 particle, and interpret any deviation from them as a signature of internal structure. 
 
\section{The $\gamma^* \Delta \Delta$ light-front helicity amplitudes}
\label{sec4}

In the following,  we consider the electromagnetic 
$\Delta \to \Delta$ transition when viewed from a light front moving towards 
the $\Delta$. Equivalently, this corresponds to  
a frame where the baryons have a 
large momentum-component along the $z$-axis chosen along the direction of 
$P = (p + p^\prime)/2$, where $p$ ($p^\prime$) are the initial (final) 
baryon four-momenta. We indicate the baryon light-front + component by $P^+$ 
(defining $a^\pm \equiv a^0 \pm a^3$). 
We can furthermore choose a symmetric frame where the virtual photon 
four-momentum $q$ has $q^+ = 0$, 
and has a transverse component (lying in the $xy$-plane)
indicated by the transverse vector $\vec q_\perp$, satisfying 
$q^2 = - {\vec q_\perp}^{\, 2} \equiv - Q^2$. 
In such a symmetric frame, the virtual photon only couples to forward moving 
partons and the + component of the 
electromagnetic current $J^+$ 
has the interpretation of the quark charge density operator. It is  
given by~: $J^+(0) = +2/3 \, \bar u(0) \gamma^+ u(0) - 1/3 \, 
\bar d(0) \gamma^+ d(0)$, considering only $u$ and $d$ quarks. 
Each term in the expression is a positive operator since 
$\bar q \gamma^+ q \propto | \gamma^+ q |^2$. 

We start by expressing the matrix elements of the $J^+(0)$ operator 
in the $\Delta$ as~:  
\begin{eqnarray}
\langle P^+, \frac{\vec q_\perp}{2}, \lambda^\prime | J^+(0) | 
P^+, -\frac{\vec q_\perp}{2}, \lambda  \rangle 
&=& (2 P^+)  e^{i  (\lambda - \lambda^\prime) \phi_q} 
\, A_{\lambda^\prime \, \lambda} (Q^2),
\label{eq:deldens1}
\end{eqnarray}
where $\lambda, \lambda^\prime$ denotes the $\Delta$ light-front
helicities, 
and where $\vec q_\perp = Q ( \cos \phi_q \hat e_x + \sin \phi_q \hat e_y )$. 
The helicity form factors 
$A_{\lambda^\prime \, \lambda}$ depend on $Q^2$ only and can equivalently 
be expressed in terms of $F^\ast_{1,2,3,4}$ as~:
\begin{eqnarray}
A_{\frac{3}{2}\frac{3}{2}}&=& A_{-\frac{3}{2} -\frac{3}{2}} = 
F_1^\ast - \frac{\tau}{2}\,F_3^\ast, 
\nonumber \\
A_{\frac{3}{2}\frac{1}{2}} &=& - A_{-\frac{3}{2} -\frac{1}{2}}=
A_{-\frac{1}{2} -\frac{3}{2}} = - A_{\frac{1}{2} \frac{3}{2}} 
= \frac{\tau^{1/2}}{\sqrt{3}} \left[2 F_1^\ast - F_2^\ast 
- \tau \left(F_3^\ast - \frac{1}{2}\, F_4^\ast \right)\right],
\nonumber \\
A_{\frac{3}{2} -\frac{1}{2}}&=& A_{-\frac{3}{2} \frac{1}{2}}=
A_{-\frac{1}{2} \frac{3}{2}}=A_{\frac{1}{2} -\frac{3}{2}} 
= \frac{\tau}{\sqrt{3}}\left[- 2 F_2^\ast + \frac{1}{2} \,F_3^\ast 
+ \tau \, F_4^\ast \right],
\nonumber \\
A_{\frac{3}{2} -\frac{3}{2}}&=& -A_{-\frac{3}{2} \frac{3}{2}}= 
- \frac{1}{2} \tau^{3/2} \,F_4^\ast , 
\nonumber \\
A_{\frac{1}{2}\frac{1}{2}}&=&A_{-\frac{1}{2} -\frac{1}{2}} 
= \left(1 - \frac{4}{3} \tau\right)F_1^\ast + \frac{\tau}{3} 
\left[4 F_2^\ast - \left(\frac{1}{2} - 2 \tau\right)F_3^\ast 
-2 \tau  F_4^\ast \right],
\nonumber \\
A_{\frac{1}{2} -\frac{1}{2}} &=& - A_{- \frac{1}{2} \frac{1}{2}} 
=\frac{\tau^{1/2}}{3} 
\Big[ 4 F_1^\ast - 2 \left(1 - 2 \tau \right)F_2^\ast - 2 \tau F_3^\ast 
+ \tau \left(\frac{1}{2}- 2 \tau\right)F_4^\ast \Big]. 
\label{eq:delhelff}
\end{eqnarray}
\newline
\indent
Since  the $\Delta \to \Delta$ electromagnetic transition is described by 
four independent form factors, one finds two angular conditions 
among the helicity form factors of Eq.~(\ref{eq:delhelff})~:
\begin{eqnarray}
0&=&\left(1+4\tau\right)\sqrt{3} A_{\frac{3}{2}\frac{3}{2}}- 
8 \tau^{1/2} A_{\frac{3}{2}\frac{1}{2}} 
+ 2 A_{\frac{3}{2} -\frac{1}{2}}
-\sqrt{3} A_{\frac{1}{2}\frac{1}{2}},
\nonumber \\
0&=&4\tau^{3/2} A_{\frac{3}{2}\frac{3}{2}}
+\sqrt{3} \left(1-2\tau\right) A_{\frac{3}{2}\frac{1}{2}}
+ \frac{1}{2} A_{\frac{3}{2} -\frac{3}{2}}- 
\frac{3}{2} A_{\frac{1}{2} -\frac{1}{2}}. 
\end{eqnarray}

\section{The transverse charge densities for a spin-3/2 particle}
\label{sec5}

We define a quark charge density for a spin-3/2 particle, such as the 
$\Delta(1232)$, in a state of definite light-cone helicity $\lambda$, 
by the Fourier transform~:
\begin{eqnarray}
\rho^\Delta_{\lambda} (b) 
&\equiv& \int \frac{d^2 \vec q_\perp}{(2 \pi)^2} 
\, e^{- i \, \vec q_\perp \cdot \vec b} \, \frac{1}{2 P^+} 
\langle P^+, \frac{\vec q_\perp}{2}, \lambda 
| J^+ | P^+, \frac{- \vec q_\perp}{2}, \lambda  \rangle  
\nonumber \\
&=& \int_0^\infty \frac{d Q}{2 \pi} Q \, 
J_0(Q b) \, A_{\lambda \lambda}(Q^2).
\label{eq:dens2}
\end{eqnarray}
The two independent quark charge densities for a spin-3/2 state 
of definite helicity are given by 
$\rho^\Delta_{\frac{3}{2}}(b)$ and 
$\rho^\Delta_{\frac{1}{2}}(b)$. 
Note that for a pointlike particle, the `natural' values of 
Eq.~(\ref{eq:sugravalues}) lead to $A_{\frac{3}{2} \frac{3}{2}} (Q^2) = 1$, 
implying 
\begin{eqnarray}
\rho_{\frac{3}{2}}(\vec b) = \delta^2(\vec b).  
\end{eqnarray}

The above charge densities provide us with two combinations of
the four independent $\Delta$ FFs. To get information from 
the other FFs, we 
consider the charge densities in a spin-3/2 state with transverse spin.  
We denote this transverse polarization direction by 
$\vec S_\perp = \cos \phi_S \hat e_x + \sin \phi_S \hat e_y$, and 
the $\Delta$ spin projection along the direction of $\vec S_\perp$ by 
$s_\perp$. 
We first express the transverse spin basis in terms of the helicity
basis for spin-3/2. For the states of transverse spin $s_\perp = \frac{3}{2}$ 
and $s_\perp = \frac{1}{2}$ this yields~:
\begin{eqnarray}
|s_\perp=+\frac{3}{2}\rangle&=&
\frac{1}{\sqrt{8}}\left\{e^{-i\phi_S}\,|\lambda = +\frac{3}{2}\rangle
+\sqrt{3}\,| \lambda = +\frac{1}{2}\rangle \right. \nonumber \\
&&\left. \hspace{.5cm} +\sqrt{3}\,e^{i\phi_S}\,|\lambda = -\frac{1}{2}\rangle
+e^{2i\phi_S}\,|\lambda = -\frac{3}{2}\rangle\right\},
\nonumber \\
|s_\perp=+\frac{1}{2}\rangle&=&
\frac{1}{\sqrt{8}}\left\{\sqrt{3}\,e^{-i\phi_S}\,|\lambda =+\frac{3}{2}\rangle
+|\lambda = +\frac{1}{2}\rangle \right.  \nonumber \\
&&\left. \hspace{.5cm} -e^{i\phi_S}\,|\lambda = -\frac{1}{2}\rangle
-\sqrt{3}\,e^{2i\phi_S}\,|\lambda = -\frac{3}{2}\rangle\right\},
\end{eqnarray}
where the states on the {\it rhs} are the spin-3/2 helicity 
eigenstates. 

We can then define the charge densities in a spin-3/2 state 
with transverse spin $s_\perp$ as~: 
\begin{eqnarray}
\rho^\Delta_{T \, s_\perp}(\vec b) 
&\equiv& \int \frac{d^2 \vec q_\perp}{(2 \pi)^2} \,
e^{- i \, \vec q_\perp \cdot \vec b} \, \frac{1}{2 P^+} 
\langle P^+, \frac{\vec q_\perp}{2}, s_\perp  
\,|\, J^+(0) \,|\, P^+, -\frac{\vec q_\perp}{2}, s_\perp   \rangle. 
\label{eq:dens3}
\end{eqnarray}
By working out the Fourier transform in Eq.~(\ref{eq:dens3}) 
for the two cases where $s_\perp = \frac{3}{2}$ 
and $s_\perp = \frac{1}{2}$, using the $\Delta$ helicity form factors 
of Eq.~(\ref{eq:delhelff}), one obtains~: 
\begin{eqnarray}
\rho^{\Delta}_{T\, \frac{3}{2}}(\vec b)
=\int_0^{+\infty}\frac{d Q}{2\pi}\,Q\,
&\Big[& J_0(Qb) \, \frac{1}{4} \left(A_{\frac{3}{2} \frac{3}{2}} 
+ 3 A_{\frac{1}{2} \frac{1}{2}} \right) \nonumber \\
&-& \sin(\phi_b-\phi_S) \, J_1(Qb) \,
\frac{1}{4} \left(2 \sqrt{3} A_{\frac{3}{2} \frac{1}{2}} 
+ 3 A_{\frac{1}{2} -\frac{1}{2}} \right) 
\nonumber \\
&-&\cos[2(\phi_b-\phi_S)] \, J_2(Qb) \,
\frac{\sqrt{3}}{2} A_{\frac{3}{2} -\frac{1}{2}} \nonumber \\
&+& \sin[3(\phi_b-\phi_S)] \, J_3(Qb) \,
\frac{1}{4} A_{\frac{3}{2} -\frac{3}{2}}
\Big], 
\label{eq:dens4} 
\end{eqnarray}
and
\begin{eqnarray}
\rho^{\Delta}_{T\, \frac{1}{2}}(\vec b)
=\int_0^{+\infty}\frac{d Q}{2\pi}\,Q\,
&\Big[& J_0(Qb) \, \frac{1}{4} \left(3 A_{\frac{3}{2} \frac{3}{2}} 
+ A_{\frac{1}{2} \frac{1}{2}} \right) \nonumber \\
&-& \sin(\phi_b-\phi_S) \, J_1(Qb) \, 
\frac{1}{4} \left(2 \sqrt{3} A_{\frac{3}{2} \frac{1}{2}} 
- A_{\frac{1}{2} -\frac{1}{2}} \right) 
\nonumber \\
&+&\cos[2(\phi_b-\phi_S)] \, J_2(Qb) \, 
\frac{\sqrt{3}}{2} A_{\frac{3}{2} -\frac{1}{2}}  \nonumber \\
&-&\sin[3(\phi_b-\phi_S)] \, J_3(Qb) \, 
\frac{3}{4} A_{\frac{3}{2} -\frac{3}{2}}
\Big],
\label{eq:dens5}
\end{eqnarray}
where we defined the angle $\phi_b$ in the transverse plane as, 
$\vec b  = b ( \cos \phi_b \hat e_x + \sin \phi_b \hat e_y )$. 
One notices from Eqs.~(\ref{eq:dens4},\ref{eq:dens5}) 
that the transverse charge densities display monopole, dipole, 
quadrupole, and octupole field patterns, which  are determined   
by the helicity form factors with zero, one, two, or three 
units of helicity flip respectively between the initial and final $\Delta$ states.  

It is instructive to evaluate the electric dipole moment (EDM)  
corresponding to the transverse charge densities 
$\rho^\Delta_{T \, s_\perp}$, 
which is defined as~: 
\begin{eqnarray}
\vec d^\Delta_{s_\perp} \equiv e \int d^2 \vec b \, \vec b \, 
\rho^\Delta_{T \,  s_\perp}(\vec b). 
\end{eqnarray}
Eqs.~(\ref{eq:dens4},\ref{eq:dens5}) yield~:
\begin{eqnarray}
\vec d^\Delta_{\frac{3}{2}} = 3 \, \vec d^\Delta_{\frac{1}{2}}
= - \left( \vec S_\perp \times \hat e_z \right) 
\, \left\{ G_{M1}(0) - 3 e_\Delta \right\} \, 
\left( \frac{e}{2 M_\Delta} \right). 
\label{eq:edm}
\end{eqnarray}
Expressing the spin-3/2 magnetic moment in terms of the $g$-factor
gives $G_{M1}(0) = g \frac{3}{2} e_\Delta$, so that the induced EDM  
$\vec d^\Delta_{s_\perp}$ 
is proportional to $g - 2$. The same result was found before   
for the case of spin-1/2 particles in~\cite{Carlson:2007xd} and spin-1 
particles in Ref.~\cite{Carlson:2008zc}. 
One thus observes as a universal feature that for a particle 
without internal structure (corresponding with 
$g = 2$~\cite{Ferrara:1992yc,Holstein:2006wi}), 
there is no induced EDM. 

We next evaluate the electric quadrupole moment corresponding 
to the transverse charge densities $\rho^\Delta_{T \, s_\perp}$.  
Choosing $\vec S_\perp = \hat e_x$, the electric quadrupole 
moment can be defined as~: 
\begin{eqnarray}
Q^\Delta_{s_\perp} \equiv e \int d^2 \vec b \, (b_x^2 - b_y^2) \, 
\rho^\Delta_{T \,  s_\perp}(\vec b). 
\end{eqnarray} 
From Eqs.~(\ref{eq:dens4},\ref{eq:dens5}) 
one obtains~:
\begin{eqnarray}
Q^\Delta_{\frac{3}{2}} = - Q^\Delta_{\frac{1}{2}}
&=& \frac{1}{2} 
\, \left\{ 2 \left[ G_{M1}(0) - 3 e_\Delta \right] + 
\left[ G_{E2}(0) + 3 e_\Delta \right] \right\} 
\, \left( \frac{e}{M_\Delta^2} \right). 
\label{eq:quadrup}
\end{eqnarray}
We note that for a spin-3/2 particle without internal structure, 
for which $G_{M1}(0) = 3 e_\Delta$ and $G_{E2}(0) = -3 e_\Delta$ 
according to Eq.~(\ref{eq:natvalues}), the quadrupole moment of the 
transverse charge densities vanishes.  
It is thus interesting to observe from Eq.~(\ref{eq:quadrup}) 
that $Q^\Delta_{s_\perp}$ is only sensitive to the anomalous parts 
of the spin-3/2 magnetic dipole and electric quadrupole moments, 
and vanishes for a particle without internal structure. The same observation 
was made for the case of a spin-1 particle in Ref.~\cite{Carlson:2008zc}.

At this point, it is important to emphasize the difference between densities and quadrupole moments defined in the lab and infinite momentum frames, and to discuss qualitatively the difference between Eq.~(\ref{eq:ee8c}) and Eq.~(\ref{eq:quadrup}).  For QCD in the lab frame, vacuum quark and gluon condensates as well as backward quark propagators play important roles in hadron structure.  In contrast, the infinite momentum frame has neither condensates nor backward going propagators Feynman diagrams.  To boost a state from the lab to the infinite momentum frame is as complicated as solving for the state in either frame, so at best, the effect of boosts can only be described in models. Because deep inelastic scattering naturally specifies quark distributions and transverse densities in the infinite momentum frame, and because of the wealth of experimental information about parton distributions and generalized parton distributions, it is clearly valuable to develop a physical understanding of and intuition for distributions and densities in this frame, and the extraction of transverse densities from form factors can contribute substantially to this understanding and intuition.  However, at the same time, it is essential to treat densities in this frame as physically distinct from those in the lab.

In this context, it is useful to discuss three differences between the conventional quadrupole moment $Q_\Delta$ of the three-dimensional density in the lab frame, Eq.~(\ref{eq:ee8c}), and the quadrupole moment $Q_{s_\perp}^\Delta$ in the infinite momentum frame, Eq.~(\ref{eq:quadrup}). The factor $[G_{M1}(0) - 3 e_\Delta ]$ is an electric quadrupole moment induced in the moving frame due to the magnetic dipole moment, which is obviously absent in the lab.
 As discussed previously, the combination $[ G_{E2}(0) + 3 e_\Delta]$ in Eq.~(\ref{eq:quadrup}) arises in the infinite momentum frame to describe non-point-like structure because the value for a point particle is  $- 3 e_\Delta$, whereas no such term is pertinent in the lab.  
 
Finally, the overall factor $1/2$ in Eq.~(\ref{eq:quadrup}) can be understood qualitatively by ignoring the distinction between the densities in the lab and infinite momentum frame, and considering a hypothetical limit in which one has a  3-dimensional density that is axially symmetric around the $x$ axis and the transverse two-dimensional density is just the 2-dimensional projection of this density, without any change of physics between the lab and infinite momentum frames. Taking the spin axis along the $x$ axis,  the quadrupole moment $Q_{3d}$ of the 3-dimensional distribution is defined as~:
\begin{eqnarray}
Q_{3d} &\equiv& \int dx dy dz \, ( 3 x^2 - r^2) \, \rho_{3d}(x,y,z), 
\nonumber \\
&=& \int dx dy dz \, \left[ (x^2 - y^2) + (x^2 - z^2) \right] \, 
\rho_{3d}(x,y,z) .
\label{eq:q3d}
\end{eqnarray}
For a 3-dimensional charge distribution that is invariant under rotations
around the axis of the spin, the two terms proportional to $(x^2 - y^2)$ 
and $(x^2 - z^2)$ in Eq.~(\ref{eq:q3d}) give equal contributions yielding~:
\begin{eqnarray}
Q_{3d} &=& 2 \int dx dy dz \, ( x^2 - y^2) \, \rho_{3d}(x,y,z).
\end{eqnarray}
Introducing the 2-dimensional charge density in the $xy$-plane as~:
\begin{eqnarray}
\rho_{2d}(x,y) &=& \int dz \, \rho_{3d}(x,y,z), 
\end{eqnarray}
one immediately obtains the relation 
\begin{eqnarray}
Q_{3d} = 2 \, Q_{2d} ,
\label{eq:q2d}
\end{eqnarray}
with the quadrupole moment of the 2-dimensional charge density defined as~:
\begin{eqnarray}
Q_{2d} &\equiv& \int dx dy \, ( x^2 - y^2) \, \rho_{2d}(x,y).
\end{eqnarray}
Because $Q_{3d}$ is proportional to $G_{E2}(0)$ in our hypothetical case, we see that 
Eq.~(\ref{eq:q2d}) yields a $Q_{2d}$ that is half the value of $G_{E2}(0)$, 
consistent with Eq.~(\ref{eq:quadrup}).   We will return to the differences between physics in the lab and infinite momentum frames and between Eq.~(\ref{eq:ee8c}) and Eq.~(\ref{eq:quadrup}) in the discussion of models in the final section. 

In the same way as for the quadrupole moment, we can also evaluate the electric octupole moment corresponding 
with the transverse charge densities $\rho^\Delta_{T \, s_\perp}$.  
Choosing $\vec S_\perp = \hat e_x$, the electric octupole 
moment can be defined as~: 
\begin{eqnarray}
O^\Delta_{s_\perp} &\equiv& e \int d^2 \vec b \, b^3 \, \sin (3 \phi_b) \, 
\rho^\Delta_{T \,  s_\perp}(\vec b), \nonumber \\
&=& e \int d^2 \vec b \, b_y \, (3 b_x^2 - b_y^2) \, 
\rho^\Delta_{T \,  s_\perp}(\vec b).
\end{eqnarray}
From Eqs.~(\ref{eq:dens4},\ref{eq:dens5}) 
one obtains~:
\begin{eqnarray}
O^\Delta_{\frac{3}{2}} = - \frac{1}{3} O^\Delta_{\frac{1}{2}}
&=& \frac{3}{2}
\, \Big\{ - G_{M1}(0) - G_{E2}(0) + G_{M3}(0) + e_\Delta  \Big\} 
\, \left( \frac{e}{2 M_\Delta^3} \right). 
\label{eq:octupole}
\end{eqnarray}
We note that for a spin-3/2 particle without internal structure, 
for which $G_{M1}(0) = 3 e_\Delta$, $G_{E2}(0) = -3 e_\Delta$,  
and $G_{M3}(0) = - e_\Delta$ according to Eq.~(\ref{eq:natvalues}),  
the electric octupole moment of the transverse charge densities 
vanishes.

\section{Lattice calculation for the $\Delta(1232)$ e.m. form factors}
\label{sec6}

Since the $\Delta(1232)$ e.m. FFs are not known 
experimentally, apart from the scarce phenomenological information described 
in Section~\ref{sec2}, 
we will rely on recent lattice QCD 
calculations~\cite{Alexandrou:2007we} for these FFs.  
For the $\Delta$ magnetic dipole moment, first dynamical results,  
using a background field method, with $N_F=2+1$ quark flavors
were presented in Ref.~\cite{Aubin:2008qp}. 
For the $\Delta$ e.m. FFs, first dynamical results were presented in 
Ref.~\cite{Alexandrou:2008bn}. We will discuss the latter calculations in 
more detail in the present section.  

We use three types of simulations that we refer to as SIM-I, SIM-II and
SIM-III.
In the SIM-I simulations, we use Wilson fermions and the standard Wilson 
plaquette 
gauge-action in the quenched approximation. In the SIM-II simulations, we use two dynamical degenerate flavors of Wilson fermions
and the standard Wilson plaquette 
gauge-action. Finally, in the SIM-III simulations, we use the following mixed action  scheme.  
For the dynamical sea quarks, we use the staggered Asqtad
action by the MILC collaboration~\cite{Bernard:2001MILC}
and include two degenerate flavors of light  quarks plus strange quarks. 
The strange quark mass is
fixed to the physical strange quark mass. For the valence quarks we use
domain wall fermions (DWF) that preserve a form of chiral symmetry on the 
lattice. 
The domain wall quark mass takes the
values given in Tables~\ref{Table:params} and has been
tuned by adjusting the lightest pseudoscalar meson in the Asqtad
calculation~\cite{Bernard:2001MILC} to have the same mass as the
pseudoscalar meson using domain-wall fermions. Technical details
of this tuning procedure are given in Refs.~\cite{Renner:2004,Hagler:2007xi}. 
Using this mixed action, a number of interesting observables were recently
evaluated, including the nucleon axial charge, 
where hybrid action chiral extrapolation  to the physical point has been
carried  out~\cite{Chen:2007ug}  and moments of generalized form factors, which are also chirally extrapolated\cite{Hagler:2007xi}. 

In this work we present a comparison between results with dynamical fermions
obtained using the Wilson and the domain wall
  fermion discretization schemes. For
finite lattice spacings these  
are different actions and  agreement between them provides 
a  non-trivial check of lattice cutoff effects. 
Results in the quenched approximation have the advantage that they are easily
produced and have smaller statistical errors.
Therefore they can be used to optimize the construction
of the three point functions. In addition, comparison
with the unquenched data at a similar pion mass provides a measure
of unquenching effects. 
  For all simulations considered in this work the $\Delta$
is a stable particle.
On the lattice we calculate the form factors $G_{E0}, G_{E2}, G_{M1}$ 
and $G_{M3}$ of which the electric
charge, $G_{E0}$, and the magnetic dipole, $G_{M1}$, form factors are dominant.

\subsection{Ground state dominance}
We use an interpolating field with the quantum numbers of the
$\Delta^+$ baryon to create an initial state from the vacuum with non-zero
overlap with the  $\Delta^+$-state. The standard interpolating field
that reduces to the quark-model wave function in the non-relativistic
limit is given by
\begin{equation}
   {\bf\chi}^{\Delta^+}_{\sigma\alpha}(x) = \frac{1}{\sqrt{3}} \epsilon^{abc}\Bigl[2\left({\bf u}^{a\top}(x) C\gamma_\sigma {\bf d}^b(x)\right)
 {\bf u}_\alpha^c(x)  
+ \left({\bf u}^{a\top}(x) C\gamma_\sigma {\bf u}^b(x)\right) {\bf d}_\alpha^c(x)\Bigr]\, ,
\eqlab{interpolatingfield}
\end{equation}
where $C$ is the charge conjugation matrix.
We note that this interpolating field creates both a spin-$3/2$ and
 a spin-$1/2$ state. However the overlap of this interpolating
field with the spin-$1/2$ $\Delta^+$ that is higher in energy is very much
suppressed and has negligible contribution to the correlator 
functions~\cite{Alexandrou:2008tn}. The overlap of the interpolating field
with the spin-$3/2$ $\Delta^+$ is
\begin{equation}
   \langle \Omega| \chi_{\sigma\alpha}(0) |\Delta(p,s)\rangle = Z\, u_{\sigma\alpha}(p,s)\, , \qquad \qquad   
   \langle \Delta(p,s) |\bar \chi_{\sigma\alpha}(0) |\Omega\rangle = Z^*\, \bar u_{\sigma\alpha}(p,s)\, ,
   \eqlab{overlap}
\end{equation}
and as in Eq.~\eqref{eq:gadeldeltree} $u_\sigma$ denotes a 
Rarita-Schwinger spinor. Every 
vector component of this vector-spinor solves the free Dirac equation
\begin{equation}
   \left[ \slashed{p} - M_\Delta \right] u_\sigma(p,s) = 0\, ,
   \eqlab{freeDiracEq}
\end{equation}
and in addition the two auxiliary conditions 
\begin{equation}
   p_\sigma u^\sigma(p,s) = 0 \qquad {\rm and} \qquad \gamma_\sigma u^\sigma(p,s) = 0 
   \eqlab{RSauxiliary}
\end{equation}
are satisfied. The well known expression for the spin-sum is given by
\begin{eqnarray}
\Lambda_{\sigma \tau} & \equiv & 
\sum_{s=-3/2}^{3/2} u_\sigma(p,s) \bar u_\tau(p,s) \nonumber \\
   &=& -\frac{\slashed p + M_\Delta}{2M_\Delta}\left(g_{\sigma\tau}-\frac{\gamma_\sigma\gamma_\tau}{3} 
                                   - \frac{2p_\sigma
				     p_\tau}{3M_\Delta^2}+\frac{p_\sigma
				     \gamma_\tau-p_\tau
				     \gamma_\sigma}{3M_\Delta} \right) 
\, ,                                   
   \eqlab{RSspinsum}
\end{eqnarray}
using the normalization $\bar u^\alpha u_\alpha = -1$. 

To facilitate ground-state dominance, we employ a covariant Gaussian 
smearing~\cite{Alexandrou:1992ti} on the
quark-fields entering \Eqref{interpolatingfield}
\begin{eqnarray}
        {{\bf q}}_\beta(t,\vec x) &=& \sum_{ \vec y} [\eins + \alpha H(\vec x,\vec y; U)]^n \ q_\beta(t,\vec y) \\
	H(\vec x, \vec y; U)  &=& \sum_{\mu=1}^3 \left(U_\mu(\vec x,t)\delta_{\vec x, \vec y - \hat \mu} + U^\dagger_\mu(\vec x-\hat \mu, t) \delta_{\vec x,\vec y+\hat\mu} \right)
\end{eqnarray}
Here $q$ is the local  $u$ or $d$ quark field, ${\bf q}$ 
is the smeared quark field and $U_\mu$ 
is the $SU(3)$-gauge field. 
For the lattice spacings and pion masses considered in this work, the values
$\alpha=4.0$ and $n=50$ ensure ground state dominance with the shortest
time evolution that could  be achieved.

\subsection{Correlation functions}
While the Breit frame is suitable for the introduction of multipole form factors,
and the infinite momentum frame allows for a clean definition of transverse charge
densities, we find that the most convenient kinematical setup for the lattice calculation
is one where the final $\Delta$-state is at rest ($\vec p_f =\vec 0$).
Furthermore, since lattice calculations are carried out in a Euclidean space-time, 
for the remainder of this chapter, all expressions are given 
with Euclidean conventions~\cite{Montvay}.

We consider two-point and three-point functions constructed from the interpolating
fields defined in ~\Eqref{interpolatingfield}\, ,
\begin{eqnarray}
   G_{\sigma \tau}(\Gamma^\nu,\vec p, t) &=&\sum_{\vec x_f} e^{-i\vec x_f \cdot \vec p}\, 
   \Gamma^\nu_{\alpha'\alpha}\, \left\langle {\mathbf \chi}_{\sigma\alpha}(t,\vec x_f) \bar{\mathbf \chi}_{\tau\alpha'}(0, \vec 0) \right\rangle \eqlab{twopoint}\, , \\
   G_{\sigma\mu \tau}(\Gamma^\nu,\vec q, t) &=& \sum_{\vec x,\, \vec x_f} e^{i\vec x \cdot \vec q}\,
   \Gamma^\nu_{\alpha'\alpha}\, \left\langle {\mathbf \chi}_{\sigma\alpha}(t_f,\vec x_f) V_\mu(t,\vec x) \bar{\mathbf \chi}_{\tau\alpha'}(0, \vec 0)\right\rangle \, . \eqlab{threepoint}
\end{eqnarray}          
With lattice Wilson fermions, $V_\mu$ is the symmetrized, conserved electromagnetic current given by  
\begin{eqnarray}
   V_\mu(x)  \! \! \! &=&\! \! \! \frac{2}{6} \left( j^u_\mu(x)+j^u_\mu(x-\hat \mu)\right) - \frac{1}{6} \left(j^d_\mu(x)+j^d_\mu(x-\hat \mu)\right) \hspace*{0.5cm}{\rm where}\\
   j^q_\mu(x)\! \! \! &=&\! \!  \!\bar q(x+\hat \mu) \frac{1}{2}[\gamma_\mu + \eins] U^{-1}_\mu(x) q(x) 
              +  \bar q(x)\frac{1}{2}[\gamma_\mu - \eins] U_\mu(x) q(x+\hat\mu) \, .
\end{eqnarray}
In the hybrid calculation we use the local vector current and determine the  
renormalization constant $Z_V$ by the condition $G_E(0) = 1$ (in units 
of the electric charge) for the $\Delta^+$.
The $\Gamma$ matrices are given by
\begin{equation}
   \Gamma^4 = \frac{1}{4}(\eins + \gamma^4)\, , \qquad \qquad \Gamma^k =
   \frac{i}{4}(\eins + \gamma^4)\gamma_5\gamma_k\, , \qquad k=1, 2, 3\, .
\end{equation}
By inserting into the correlation functions complete sets of
energy momentum eigenstates  
\begin{equation}
   \sum_{n,p,\xi} \frac{M_n}{V\,E_{n(p)}} | n(p,\xi)\rangle \langle n(p,\xi) | =  \eins \quad,
   \eqlab{completeSet}
\end{equation}
with $\xi$ denoting
 all other quantum numbers like spin and its projection on some axis,
one finds that the leading contributions for large Euclidean times $t$ 
and $t_f-t$ are
\begin{eqnarray}
   G_{\sigma \tau} (\Gamma^\nu,\vec p, t)         &=& \frac{M_\Delta}{E_{\Delta(p)}} \, |Z|^2\, e^{-E_{\Delta(p)}\, t}\,  
                                                      {\rm tr} \left[\Gamma^\nu \Lambda^E_{\sigma \tau}(p) \right] + {\rm excited\ states}\, ,\\
   G_{\sigma\mu\tau}(\Gamma^\nu,\vec q, t) &=&
                                                      \frac{M_\Delta}{E_{\Delta(p_i)}}\, |Z|^2 \, e^{-M_\Delta \,(t_f-t)}\,e^{-E_{\Delta(p_i)}\, t }\, 
{\rm tr}[\Gamma^\nu \Lambda^E_{\sigma \sigma'}(p_f) \Op^E_{\sigma' \mu \tau'} 
\Lambda^E_{\tau' \tau}(p_i)] \nonumber \\
&+& {\rm excited\ states} \, .
\end{eqnarray}
To calculate the spin traces, the Euclidean  versions of \Eqref{gadeldeltree} and \Eqref{RSspinsum}
are employed.
The unknown overlap $Z$-factors and the leading time dependence cancel in 
certain ratios of three- and two- point functions. Our preferred choice is the ratio
\begin{equation}
        R_{\sigma\mu\tau}(\Gamma,\vec q,t) = \frac{G_{\sigma\mu\tau}(\Gamma,\vec q,t)}{G_{k k}(\Gamma^4,\vec 0, t_f)}\ 
				         \sqrt{\frac{G_{kk}(\Gamma^4,\vec p_i, t_f-t)G_{kk}(\Gamma^4,\vec 0  ,t)G_{kk}(\Gamma^4,\vec 0,t_f)}
					            {G_{kk}(\Gamma^4,\vec 0, t_f-t)G_{kk}(\Gamma^4,\vec p_i,t)G_{kk}(\Gamma^4,\vec p_i,t_f)}}\, ,
\label{eq:ratio}
\end{equation}
with implicit summations over the indices $k$ with $k=1,2, 3$.
It becomes time independent for large Euclidean time separations $t_f-t$ 
and $t$: 
\begin{equation}
   R_{\sigma\mu \tau}(\Gamma,\vec q,t) \to \Pi_{\sigma\mu \tau}(\Gamma,\vec q) = 
    C\ {\rm tr}\left[\Gamma\, \Lambda_{\sigma\sigma'}(p_f) \Op_{{\sigma'}\mu{\tau'}} \Lambda_{\tau'\tau}(p_i) \right] \, ,
\end{equation}
with 
\begin{equation}
C\equiv\sqrt{\frac{3}{2}}\left[\frac{2 E_{\Delta(p_i)}}{M_\Delta} 
                          +\frac{2 E^2_{\Delta(p_i)}}{M^2_\Delta} 
                          +\frac{  E^3_{\Delta(p_i)}}{M^3_\Delta} 
                          +\frac{  E^4_{\Delta(p_i)}}{M^4_\Delta} \right]^{-\frac{1}{2}},
\end{equation} 
stemming from the state normalization and the 2-point function traces.

Since we are evaluating the correlator of \Eqref{threepoint} using  
sequential inversions through the sink~\cite{Dolgov:2002zm}, 
a separate set of inversions is necessary for every choice of vector and Dirac-indices.
The total of $256$ combinations is beyond our computational resources, and hence we concentrate on a few 
carefully chosen combinations given below:
\begin{eqnarray}
   \Pi_\mu^{(1)}(\vec q) &=& \sum \limits_{j,k,l=1}^3 \epsilon_{jkl}\Pi_{j\mu
   k}(\Gamma^4, \vec q) \nonumber \\
   &=& G_{M1}\
   \frac{5i(E_{\Delta}+M_\Delta)C}{18M_\Delta^2}\left[\delta_{1,\mu}(q_3-q_2)
   + \delta_{2,\mu}(q_1-q_3) + \delta_{3,\mu}(q_2-q_1)\right] ,
\nonumber \\
\eqlab{comb1} \\
   \Pi_\mu^{(2)}(\vec q) &=& \sum \limits_{k=1}^3 \Pi_{k\mu k}(\Gamma^4, \vec
   q) \nonumber \\
   &=&   -G_{E0}\ \frac{(E_{\Delta}+2M_\Delta)C}{3M_\Delta^2}   \left[(M_\Delta+E_\Delta) \delta_{4,\mu}+iq_\mu(1-\delta_{4,\mu}) \right]\nonumber \\
   &-&G_{E2}\
   \frac{(E_{\Delta}-M_\Delta)^2C}{9M_\Delta^3}\left[(M_\Delta+E_\Delta)
   \delta_{4,\mu}+iq_\mu(1-\delta_{4,\mu}) \right] ,
\eqlab{comb2} 
\end{eqnarray}
\begin{eqnarray}
   \Pi_\mu^{(3)}(\vec q) &=& \sum \limits_{j,k,l=1}^3 \epsilon_{jkl}\Pi_{j\mu
   k}(\Gamma^j, \vec q) \nonumber \\
   &=& G_{E2}\ \frac{-iC}{3M_\Delta^2(E_{\Delta}+M_\Delta)} (q_1q_2 + q_2q_3 +
   q_3q_1) \nonumber \\
&&\hspace{3.5cm} \times
\left[(M_\Delta+E_\Delta)\delta_{4,\mu}+iq_\mu(1-\delta_{4,\mu}) \right] \nonumber \\
   &+& G_{M1}\ \frac{C}{6M_\Delta^2(E_\Delta+M_\Delta)} \sum_{k=1}^3 
            \delta_{k,\mu}\, q_1 q_2 q_3 \left(2-\frac{q_1+q_2+q_3-q_k}{q_k} \right) \nonumber \\
   &+& G_{M3}\ \frac{C}{30M_\Delta^3(E_\Delta+M_\Delta)}\sum_{k=1}^3
            \delta_{k,\mu}\biggl[(16E_\Delta+14M_\Delta)q_1 q_2 q_3 \nonumber \\
   &&\hspace{3.5cm} - 10M_\Delta(q_1 q_2 + q_2 q_3 + q_3 q_1) q_k 
\nonumber \\
   &&\hspace{3.5cm} -(8E_\Delta+7M_\Delta)\frac{q_1 q_2
   q_3}{q_k}(q_1+q_2+q_3-q_k)\biggr] \nonumber \, . \\
\eqlab{comb3} 
\end{eqnarray}
As expected, current conservation $q_\mu \Pi_\mu = 0$ is manifest in the right hand side
of the equations. 
From these expressions, all the multipole form factors can be extracted. For instance,~\Eqref{comb1} 
is proportional to $G_{M1}$, while~\Eqref{comb3} isolates $G_{E2}$ for $\mu=4$.
Furthermore, these combinations are optimal in the sense that
all momentum directions, each of which is statistically different, contribute
to a given $Q^2$-value. This symmetric construction yields a 
better estimator for the $\Delta$-matrix elements than methods 
where only one momentum-vector is accessible.

\begin{table}[h]
\small
\begin{center}
\begin{tabular}{ccccccc}
\hline\hline 
\multicolumn{7}{c}{Wilson fermions}\\
\hline
V& \# confs & $\kappa$ & $m_\pi$ & $m_\pi/m_\rho$ & $m_N$ & 
$m_\Delta$ \\ 
& & & (GeV)& & (GeV) & (GeV)\\ 
\hline
\multicolumn{7}{c}{SIM-I : Quenched, \quad $\beta=6.0,~~a^{-1}=2.14(6)$~GeV} \\
\hline
$32^3\times 64$& 200 & 0.1554 &0.563(4)& 0.645(9)  & 1.267(11) & 1.470(15)\\
$32^3\times 64$& 200 & 0.1558 &0.490(4)& 0.587(12) & 1.190(13) & 1.425(16)\\
$32^3\times 64$& 200 & 0.1562 &0.411(4)& 0.503(23) & 1.109(13) & 1.382(19)\\
\hline
\multicolumn{7}{c}{SIM-II: Unquenched, 
\quad $\beta=5.6,~~a^{-1} = 2.56(10)$~GeV}\\
\hline
  $24^3\times 40$&185~\cite{Orth:2005kq} 
&0.1575 &0.691(8)&0.701(9) &1.485(18) & 1.687(15)\\
  $24^3\times 40$&157~\cite{Orth:2005kq} 
&0.1580 &0.509(8)&0.566(12) &1.280(26) & 1.559(19)\\
 $24^3\times 32$& 200~\cite{Urbach:2005ji} 
& 0.15825 &0.384(8)& 0.453(27)&1.083(18) & 1.395(18)\\
\hline
\hline
\multicolumn{7}{c}{SIM-III: Mixed action} \\
\multicolumn{7}{c}{
Asqtad ($am_{\mbox{\tiny u,d/s}} = 0.01/0.05$),  
DWF ($ am_{\mbox{\tiny u,d}} = 0.0138$), $a^{-1} = 1.58(3)$~GeV}\\
\hline
    $V$ & \# confs & & $m_\pi$  & $m_\pi / m_\rho$ &
    $m_N$  & $m_\Delta$ \\ 
     &  & & (GeV) & &
    (GeV) & (GeV) \\ 
\hline
$28^3\times 64$ &300 &  & 0.353(2) & 0.368(8)&1.191(19) &
 1.533(27)\\
\hline\hline
\end{tabular}
\end{center}
\caption{Parameters used
  in the calculation of the form factors.}
\label{Table:params}
\end{table}

\subsection{Sink-source separation}

\begin{figure}[h]
\includegraphics[width=0.49 \linewidth]{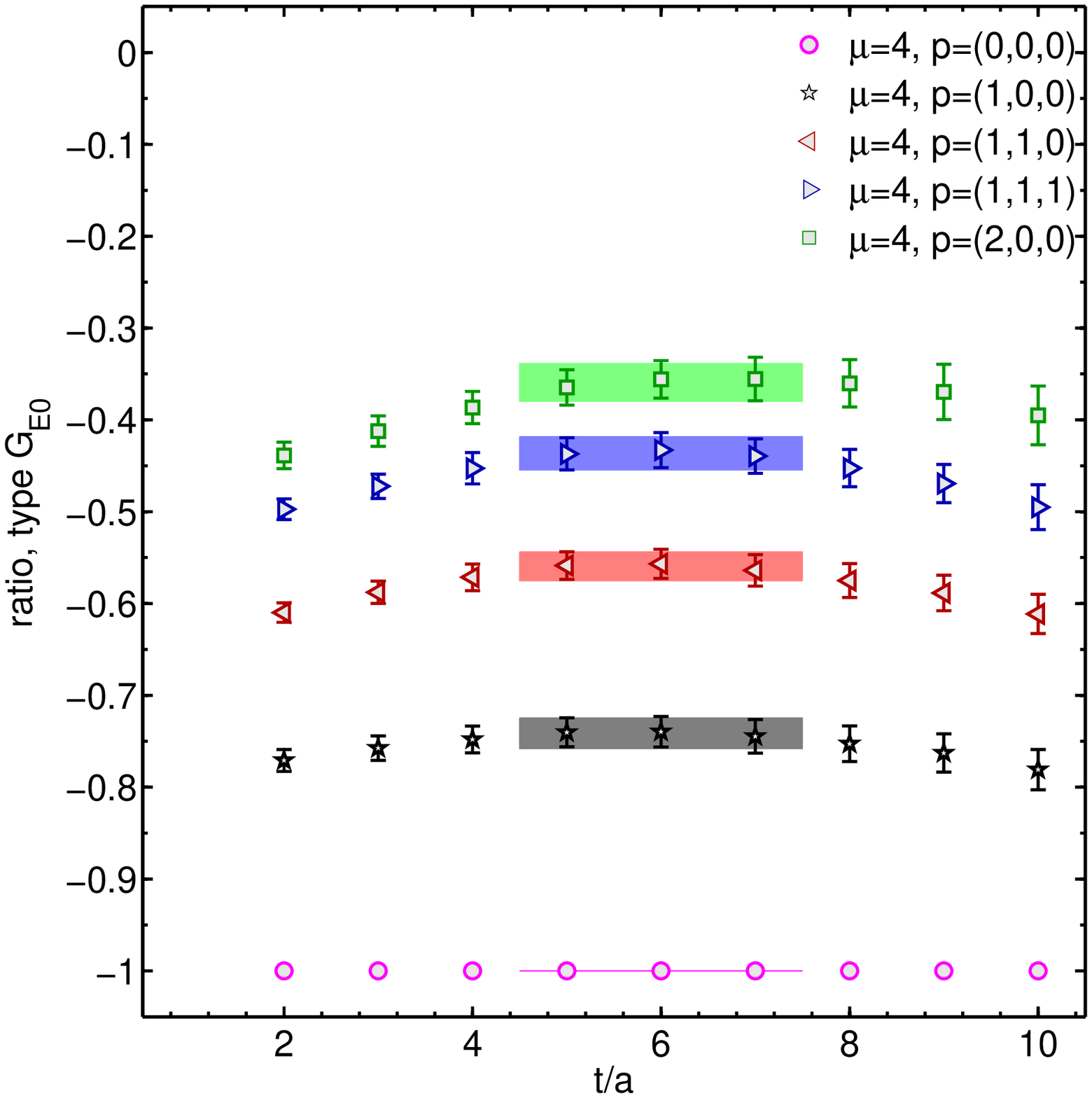}
\includegraphics[width=0.49 \linewidth]{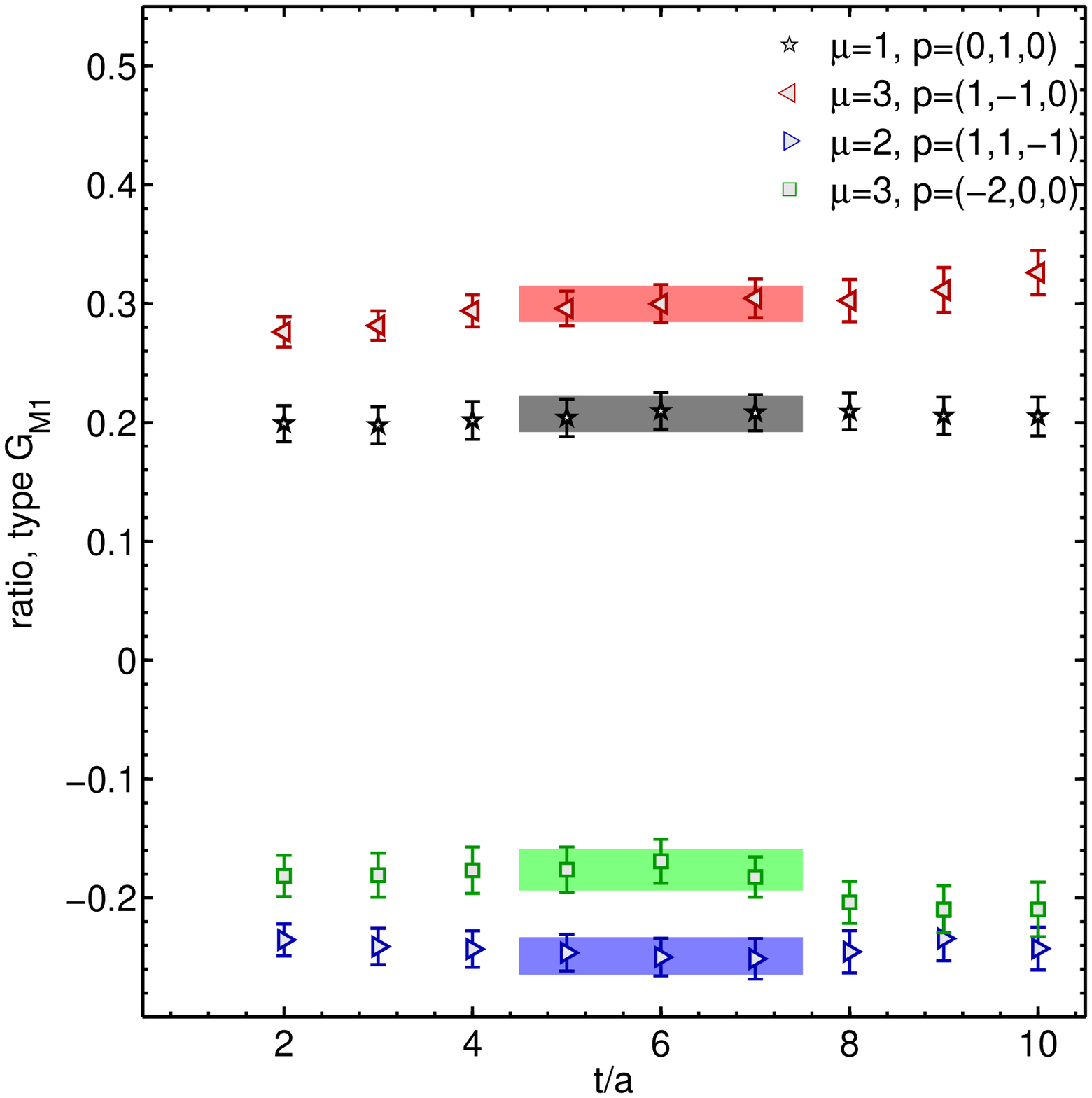}
\caption{The ratio defined in Eq.~(\ref{eq:ratio})  for a few of the 
lowest momenta. The left panel shows 
the combination from which $G_{E0}$ is extracted, whereas the right  shows 
the one that isolates $G_{M1}$. This is for the quenched case where 
due to the high statistical accuracy excited state contributions are 
most visible. For higher momenta, as well as  for the suppressed form factors 
the ratio of systematic to statistical error is lower.}
\figlab{quenched plateaus}
\end{figure}

\begin{figure}[h]
\includegraphics[width=0.49 \linewidth]{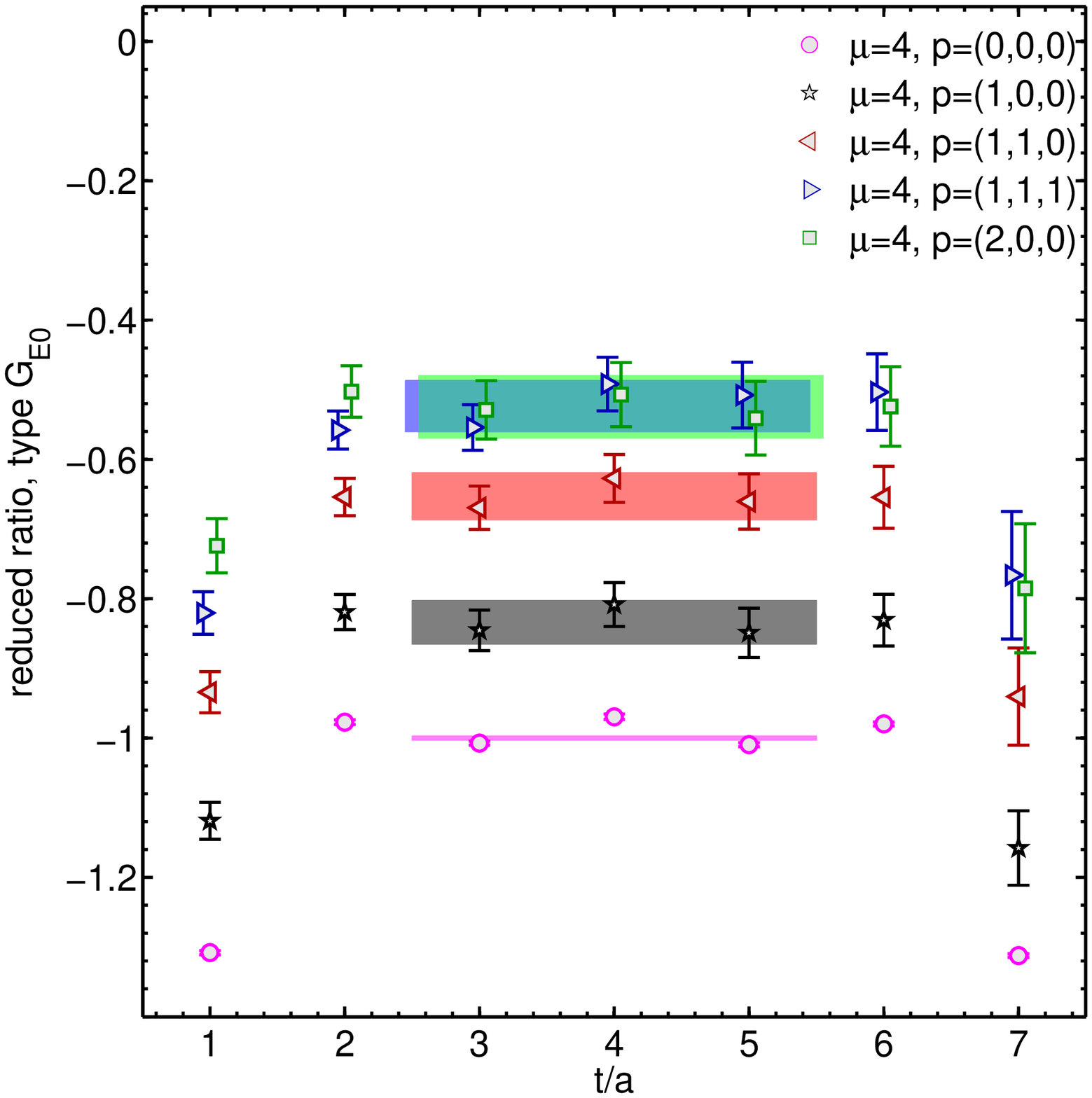}
\includegraphics[width=0.49 \linewidth]{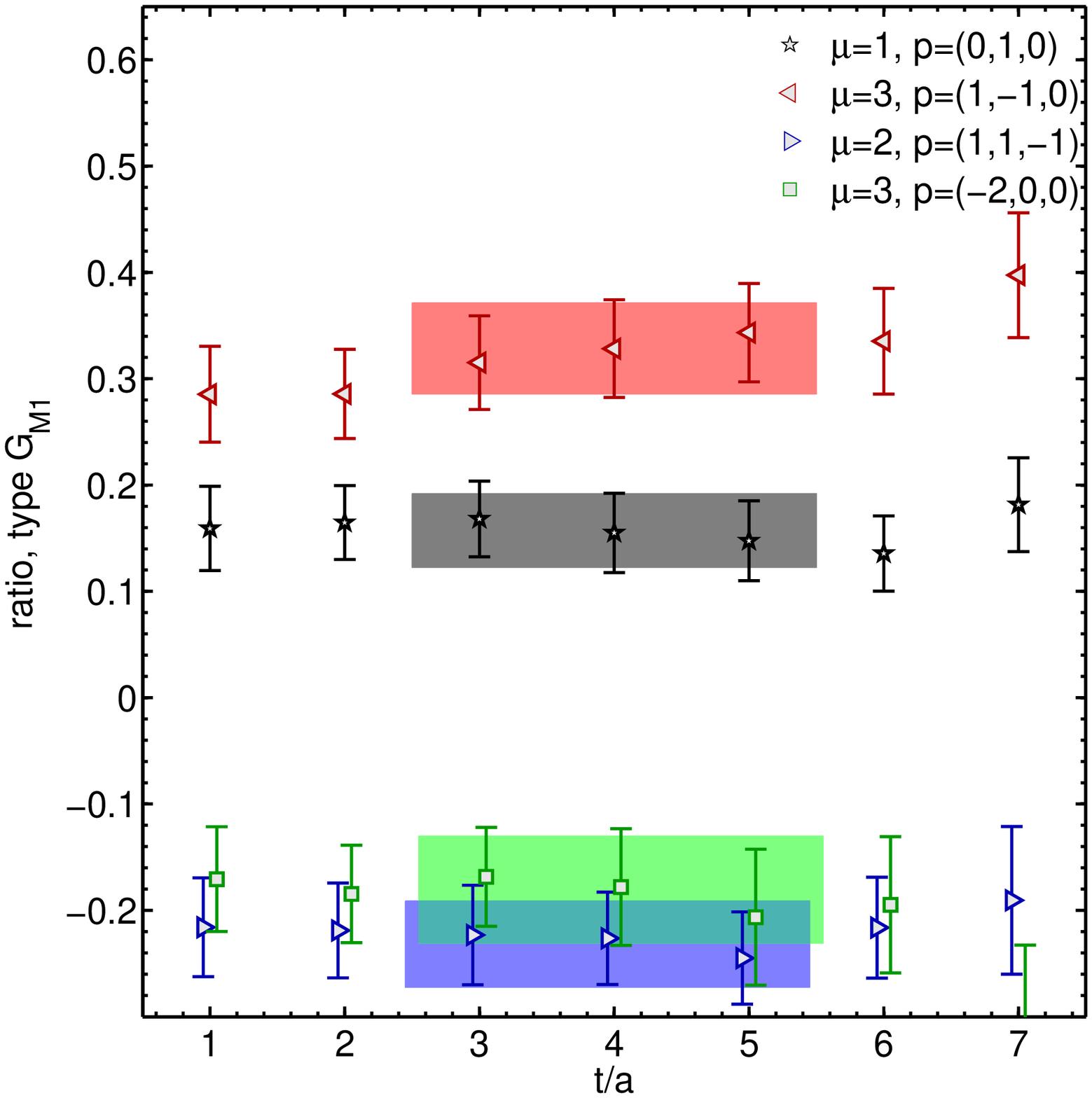}
\caption{The same as in Fig.~\ref{fig:quenched plateaus}
but for the hybrid case.}
\figlab{hybrid plateaus}
\end{figure}

Apart from the simulations themselves, sequential inversions 
are the most time consuming part
of the calculation. They are done by constructing an appropriate
sequential source,
which requires that we fix the initial and final hadron states
as well as the sink-source time separation, $t_f$. Changing $t_f$
requires a new set of sequential inversions. Therefore
we must first determine the optimal value of $t_f$. The criterion is
to choose $t_f$ as small as possible so that statistical errors 
due to the exponential decrease of the signal are minimized but large
enough so that we ensure that excited states
with $\Delta$ quantum numbers are suppressed 
when the photon couples to a quark at time $t$ from the source. 
Smearing techniques are essential for attaining early ground state
dominance as discussed in the previous subsection.
The optimal source-sink separation for a very similar
calculation to ours was determined in Refs.~\cite{Alexandrou:2007dt,Alexandrou:2007we}.
The authors find that a distance of about 1~fm is
sufficient to suppress excited state contributions
well below our targeted statistical errors.
This translates into a separation
of 12 time slices for SIM-I, 13 for SIM-II and 8 for SIM-III.
That this is indeed sufficient, is demonstrated in 
\Figref{quenched plateaus} and  \Figref{hybrid plateaus} where we show a few 
representative
 ratios with plateaus for SIM-I and SIM-III respectively.

\subsection{Volume dependence}
Another potential source of  systematic error
is the spatial size of our lattices.
Given that for the quenched case we use a lattice of spatial size about 3~fm,
we  expect finite volume effects to be negligible.
Similarly, for the mixed action,
 the spatial lattice size is $L_s=3.5$~fm
 giving $L_s m_\pi=6.4$ ensuring small finite volume effects.
In Ref.~\cite{Alexandrou:2007dt} it was in fact shown that for the N to $\Delta$
matrix elements,   finite volume effects
are small for  $L_sm_\pi\stackrel{>}{\sim}4.5$. 
Except for the lightest mass for dynamical
Wilson fermions for which $L_sm_\pi\sim3.6$,  for all our
quark masses we have
 $L_sm_\pi>4.6$, and therefore we expect finite volume effects to be small.

\subsection{Data analysis}
For a given value of $Q^2$, the combinations given in Eqs.~\eref{comb1} to~\eref{comb3} are evaluated
for all different directions of $\vec q$ resulting in this $Q^2$, as well as for 
all four directions $\mu$ of the current. This leads to an over-constrained linear
system of equations, which is then solved in the least-squares sense yielding 
estimates of $G_{E0}$, $G_{E2}$, $G_{M1}$ and $G_{M3}$. This estimation is
embedded into a jackknife binning procedure, 
thus providing  statistical 
errors for the form factors that take all correlation and 
autocorrelation effects
into account. 
\begin{figure}[h]
\begin{center}
\includegraphics[width =9.5cm]{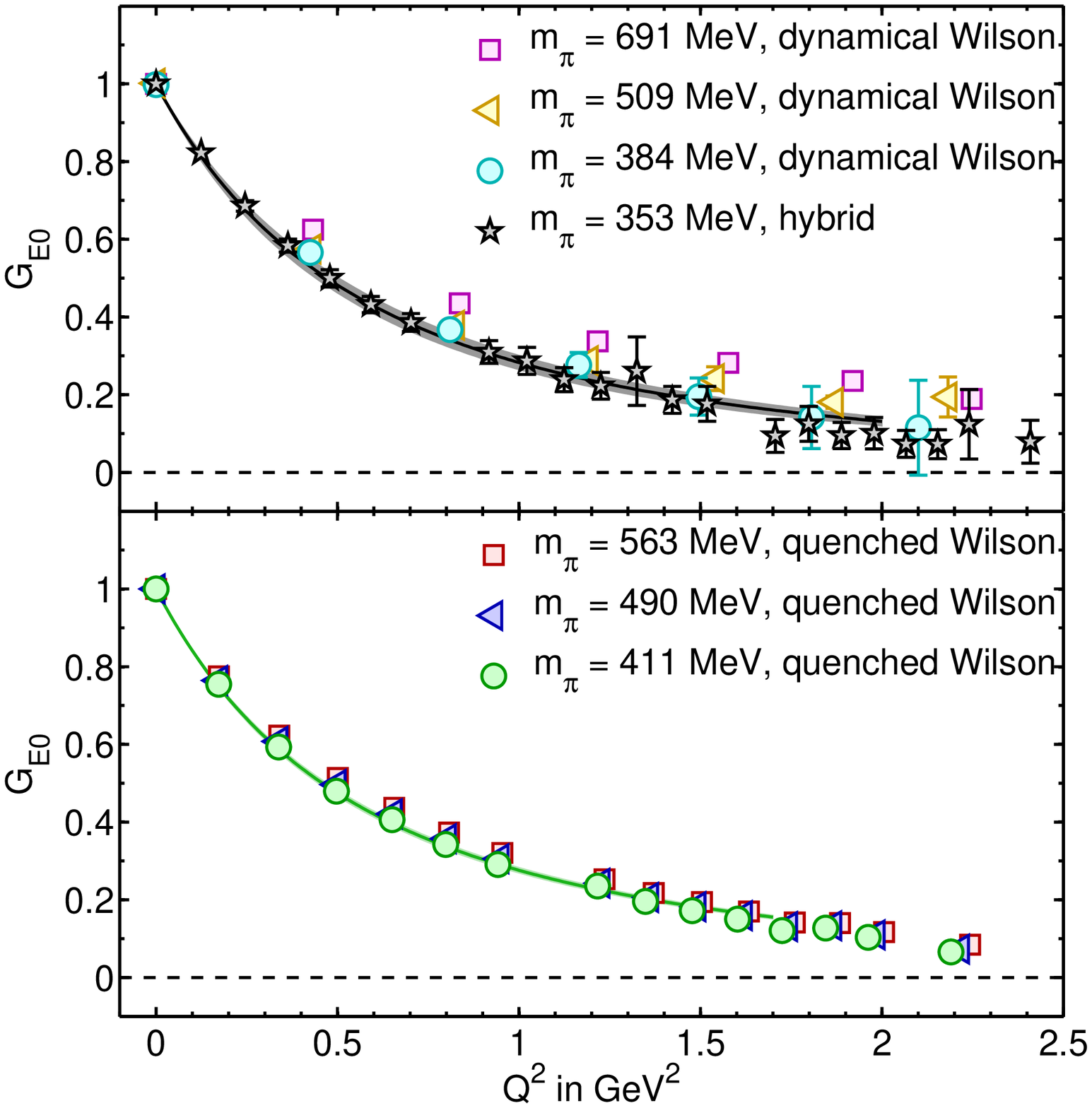}
\end{center}
\caption{
Comparison of three different QCD lattice calculations for the 
$\Delta^+(1232)$  form factor
 $G_{E0}$. The upper curve is for SIM-II and SIM-III using dynamical quarks,
whereas the lower curve is for SIM-I using the quenched 
approximation~\cite{Alexandrou:2007we}.
The lines show the fits to a dipole form of the lattice results at the smallest
pion mass. For the case
of dynamical quarks  (upper graph) the fit is made to the results
obtained using the mixed action. 
The error band is calculated using a jackknife analysis of the fit parameters.
}
\label{fig:ge0}
\end{figure}

\begin{figure}[h]
\begin{center}
\includegraphics[width =9.5cm]{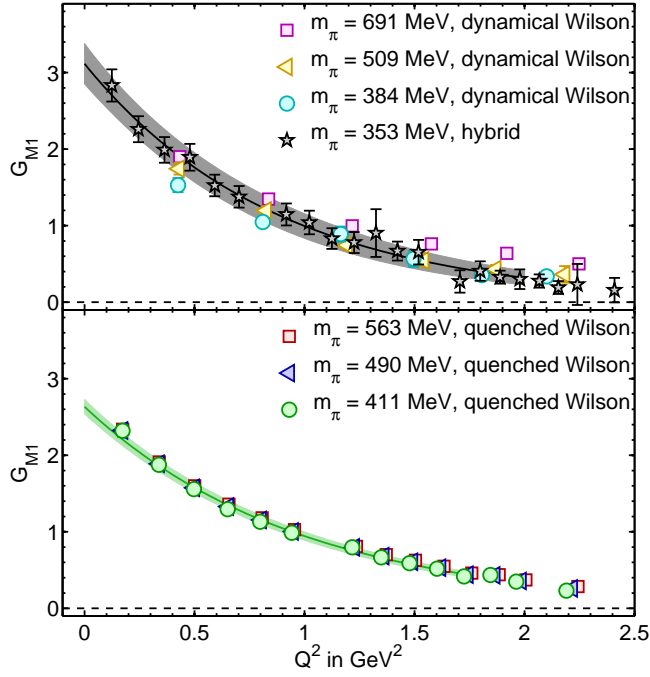}
\caption{
Comparison of three different QCD lattice calculations for the 
$\Delta^+(1232)$ magnetic dipole form factors
 $G_{M1}$. The lines show the fits to an exponential
 form of the lattice results at the smallest
pion mass. The rest of the notation is the same as that in Fig.~\ref{fig:ge0}.
}
\label{fig:gm1}
\end{center}
\end{figure}

\begin{figure}[h]
\begin{center}
\includegraphics[width =9.5cm]{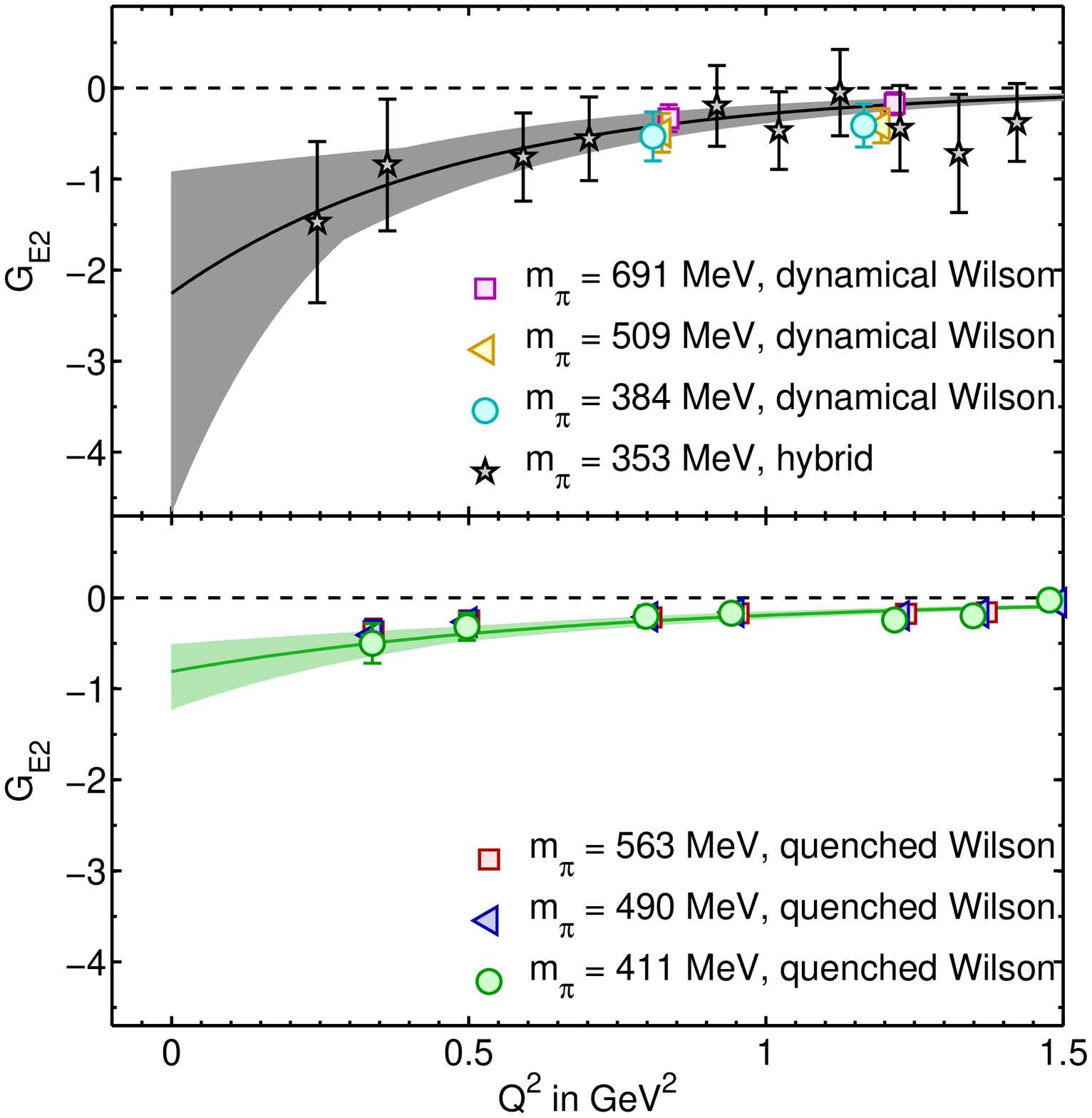}
\caption{
Comparison of three different QCD lattice calculations for the 
$\Delta^+(1232)$ electric quadrupole form factors
 $G_{E2}$.  The notation is the same as that in Fig.~\ref{fig:gm1}.
}
\label{fig:ge2}
\end{center}
\end{figure}

\subsection{Simulation parameters}
The simulation parameters of our calculation are given 
in Table~\ref{Table:params}.
In the case of Wilson fermions, the lattice spacing has been estimated
using the nucleon mass in the chiral limit~\cite{Alexandrou:2006ru}.
The value extracted using the nucleon mass 
 is in agreement with the determination
using the hadronic scale $r_0$ defined via the force between static quarks
at intermediate distance~\cite{Sommer:1993ce}.
The lattice spacing in the case of staggered
fermions has been determined from
heavy-quark spectroscopy~\cite{Aubin:2004wf} as $a=0.1241$ fm 
with a statistical
uncertainty of $2\%$. 
The calculation has been performed on  relatively large
lattices 
in the quenched approximation and
 in the case of the mixed action of spatial extent 2.9~fm and 
3.4~fm respectively.
On the other hand, in the case of dynamical Wilson fermions we use
a finer lattice with a
spatial size of only 1.9~fm.

Isospin symmetry relates  results obtained for the $\Delta^+$ to 
those for the $\Delta^{++},\, \Delta^0$ and $\Delta^-$ since they differ only
by a charge-factor.
The pion and $\Delta$ masses corresponding
to the three values of the hopping parameter, $\kappa$, considered 
in simulations SIM-I and SIM-II as well as the smallest
available pion mass in  the mixed action are
 summarized in Table~\ref{Table:params}.
The results presented in the next subsection  are obtained from the 
connected contributions to the
full electromagnetic current.
 These yield the iso-vector
form factors with the iso-vector current being
given by $V_\mu^{I} = \bar u \gamma_\mu u - \bar d \gamma_\mu d$. 
In this case the disconnected diagrams cancel.
The iso-scalar form factors 
have contributions from disconnected diagrams. 
 At present,
 it is not yet computationally feasible to calculate 
the contributions arising from disconnected diagrams since
they require the calculation of all-to-all propagators. An exact 
evaluation of all-to-all propagators requires spatial volume 
inversions for each time slice $t$ that we are interested in. Such
a computation is not feasible. Stochastic techniques 
have been developed to estimate all-to-all propagators~\cite{disconnected}.
Various improvements are being developed in order to improve 
the measurements~\cite{disconnected-improve}.
However, these disconnected 
contributions are particularly hard to calculate 
not just because they require the
all-to-all propagators but also because they are 
noise-dominated~\cite{Jansen:2008wv}.
A calculation of disconnected contributions in the case
 of the electromagnetic form 
factors have shown that these are consistent with zero
within large statistical errors~\cite{Wilcox:2000qa}.
 We expect 
that,
like in  the case of nucleon electromagnetic form factors, the
disconnected contributions to the electromagnetic $\Delta$ form factors 
are small. Therefore assuming that the  disconnected diagrams are small
 then we obtain the  electromgnetic form factors of $\Delta$.

\begin{table}[h]
\small
\begin{center}
\begin{tabular}{c c c c c c c }
\hline
\hline
 &$m_\pi$ & $\Lambda_{E0}^2$ & $G_{M1}(0)$ & $\Lambda_{M1}^2$ & $G_{E2}(0)$ &
$\Lambda_{E2}^2$  \\
  &(GeV) & (GeV$^2$) && (GeV$^{2}$) && (GeV$^{2}$)   \\
\hline
SIM-I 		& 
  0.563(4)   &    1.243(27) & 2.695(59) & 1.015(32) & -0.46(20)& 1.06(42)\\
  & 0.490(4)   &    1.173(28) & 2.678(72) & 0.994(36) & -0.52(26) &0.91(39)\\
  & 0.411(4)   &    1.109(29) & 2.635(94) & 0.975(23) & -0.81(29) &
0.696(200) \\
\hline
 SIM-II 	&
  0.691(8)    & 1.667(106) & 2.589(78) & 1.300(81) & -0.71(49) & 1.00(64)\\
  & 0.509(8)    & 1.318(48)  & 2.68(13)  & 0.977(74) & -1.68(88) & 0.76(30)\\
  & 0.384(8)    & 1.144(54)  & 2.35(16)  & 1.015(76) & -0.87(67)  &
1.75(1.39) \\
\hline 
\\
SIM-III	& 
0.353(2)      & 1.160(78) &  3.04(24) & 0.935(122) & $-2.06^{+1.27}_{-2.35}$ &
$0.54^{+1.69}_{-0.25}$  \\
\hline \hline
\end{tabular}
\end{center}
\caption{Parameters for the $\Delta^+(1232)$ FFs  
for three different lattice QCD calculations : 
quenched Wilson (SIM-I), dynamical $N_f = 2$ Wilson (SIM-II), and 
a hybrid calculation (SIM-III) described in the text.  
For $G_{E0}$, the dipole parameterization of Eq.~(\ref{eq:ge0lattice}) 
is used. For $G_{M1}$ and $G_{E2}$, the exponential parameterization of 
Eq.~(\ref{eq:gm1ge2lattice}) is used. The errors on the parameters are
jackknife errors.}
\label{table:latticefit1}
\end{table}


\subsection{Results}
We show in Figs.~\ref{fig:ge0} and \ref{fig:gm1}  the two
dominant Sachs form factors $G_{E0}$ and $G_{M1}$ as a function of the 
$Q^2$ for all the light quark masses.  In Fig.~\ref{fig:ge2} we 
display the $Q^2$-dependence of the subdominant Sachs form factor $G_{E2}$.
Their values are tabulated in Tables~\ref{quenched-results}, 
\ref{wilson-results} and \ref{hybrid-results} of the Appendix. 

For $G_{E0}$, we consider a 
dipole parameterization of the lattice results~:
\begin{eqnarray} 
G_{E0}(Q^2) &=& \frac{1}{\left( 1 + Q^2 / \Lambda_{E0}^2 \right)^2} , 
\label{eq:ge0lattice}
\end{eqnarray}
with $\Lambda_{E0}^2$ treated as a free parameter. 
The choice of the parameterization for the $\Delta$ multipole 
form factors given in~\Eqref{ge0lattice} 
ensures that the helicity conserving form factors
 $A_{\frac{3}{2} \frac{3}{2}}$ 
and $A_{\frac{1}{2} \frac{1}{2}}$ behave as $1 / Q^4$ for large $Q^2$. 
For $G_{M1}$ and $G_{E2}$, we instead
consider exponential parameterizations 
with two free parameters for each since 
the expected large $Q^2$-dependence for these form factors
is stronger than for a dipole~:
\begin{eqnarray}
G_{M1}(Q^2) &=& G_{M1}(0) \, e^{- Q^2 / \Lambda_{M1}^2} , 
\nonumber \\
G_{E2}(Q^2) &=& G_{E2}(0) \, e^{- Q^2 / \Lambda_{E2}^2}.  
\label{eq:gm1ge2lattice} 
\end{eqnarray}
The magnetic octupole form factor $G_{M3}$ evaluated in the
quenched approximation~\cite{Alexandrou:2007we} is found to
have a small  negative value 
 albeit with large error. The results for  $G_{M3}$ using
 dynamical Wilson fermions
obtained in this work are compatible with $G_{M3}(Q^2) \simeq 0$ within
the current statistical accuracy. We will therefore not discuss further
lattice results for  $G_{M3}$.

The values of the parameters entering the dipole 
fit of Eq.~(\ref{eq:ge0lattice}) for $G_{E0}$, and the 
exponential fits of Eq.~(\ref{eq:gm1ge2lattice}) for 
$G_{M1}$ and $G_{E2}$, fitted to the three 
different lattice calculations SIM-I, SIM-II and SIM-III
are listed in Table~\ref{table:latticefit1}.

In Figs.~\ref{fig:ge0}, \ref{fig:gm1} and \ref{fig:ge2} we show
 the  fits to  the lattice results for the 
form factors $G_{E0}, G_{M1}$ and 
$G_{E2}$ for the smallest pion mass in each of the three type
of simulations.
One sees that for $G_{E0}$ all three calculations give similar results. 
For $G_{M1}$ and more so for $G_{E2}$,  a larger spread in the resulting
fits is observed. 
For $G_{E2}(0)$ the hybrid lattice calculation yields a quadrupole moment 
consistent with the large-$N_c$ value, Eq.~(\ref{eq:qdellargenc}), whereas 
the quenched and dynamical $N_f = 2$ Wilson calculations yield 
only about half this value.  
It should, however, be noted that the present dynamical calculations 
still have substantially larger error bars than the quenched calculations.

\section{The $\Delta^+$ transverse densities}
\label{sec7}

In this section, we will use the values of the parameters determined
in the lattice fits to the $\Delta$ e.m. FFs, discussed in Section \ref{sec6}, 
to evaluate the quark transverse densities in the $\Delta^+(1232)$.

\begin{figure}[H]
\begin{center}
\includegraphics[width = 0.49 \linewidth]{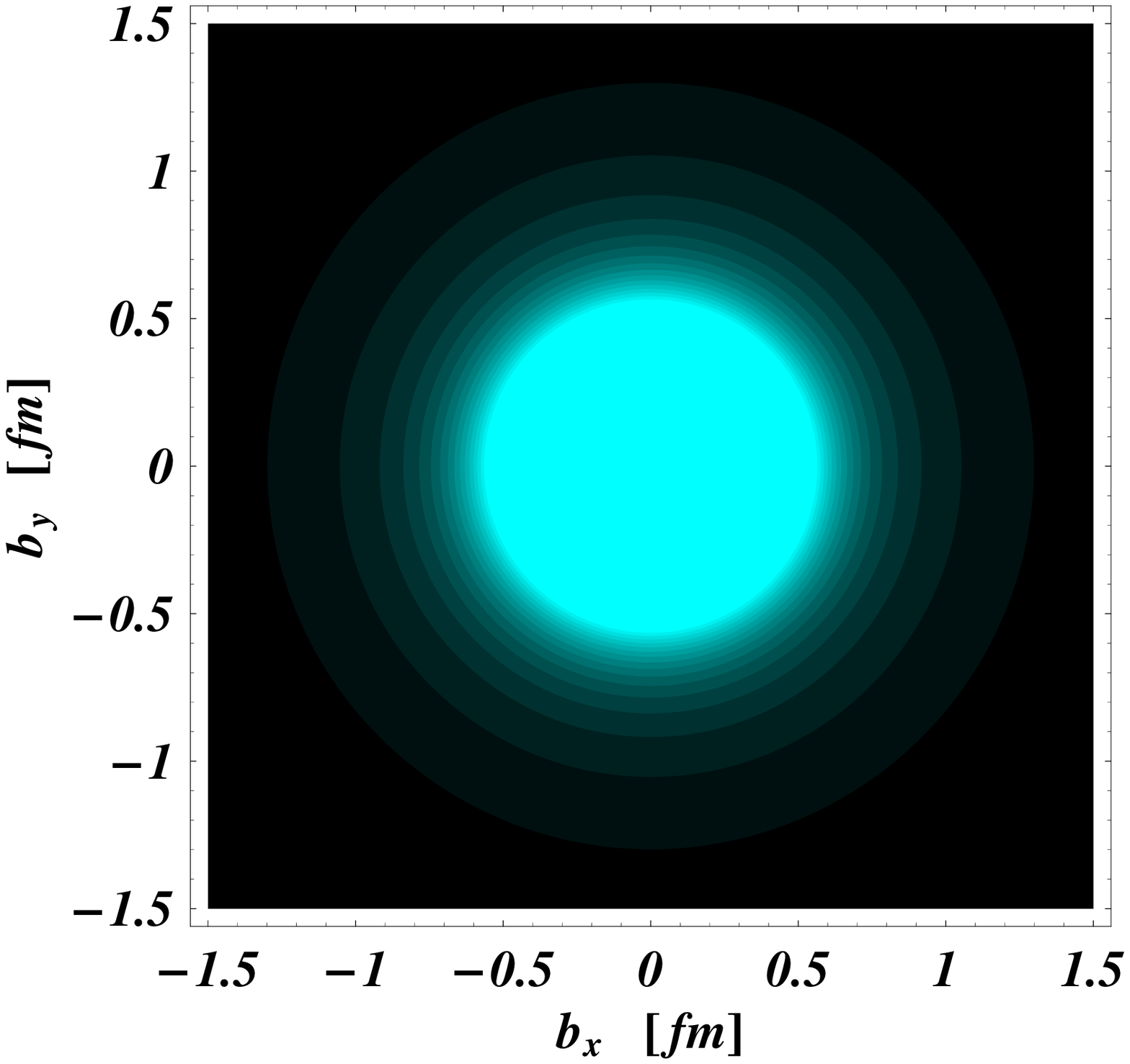}
\includegraphics[width = 0.49 \linewidth]{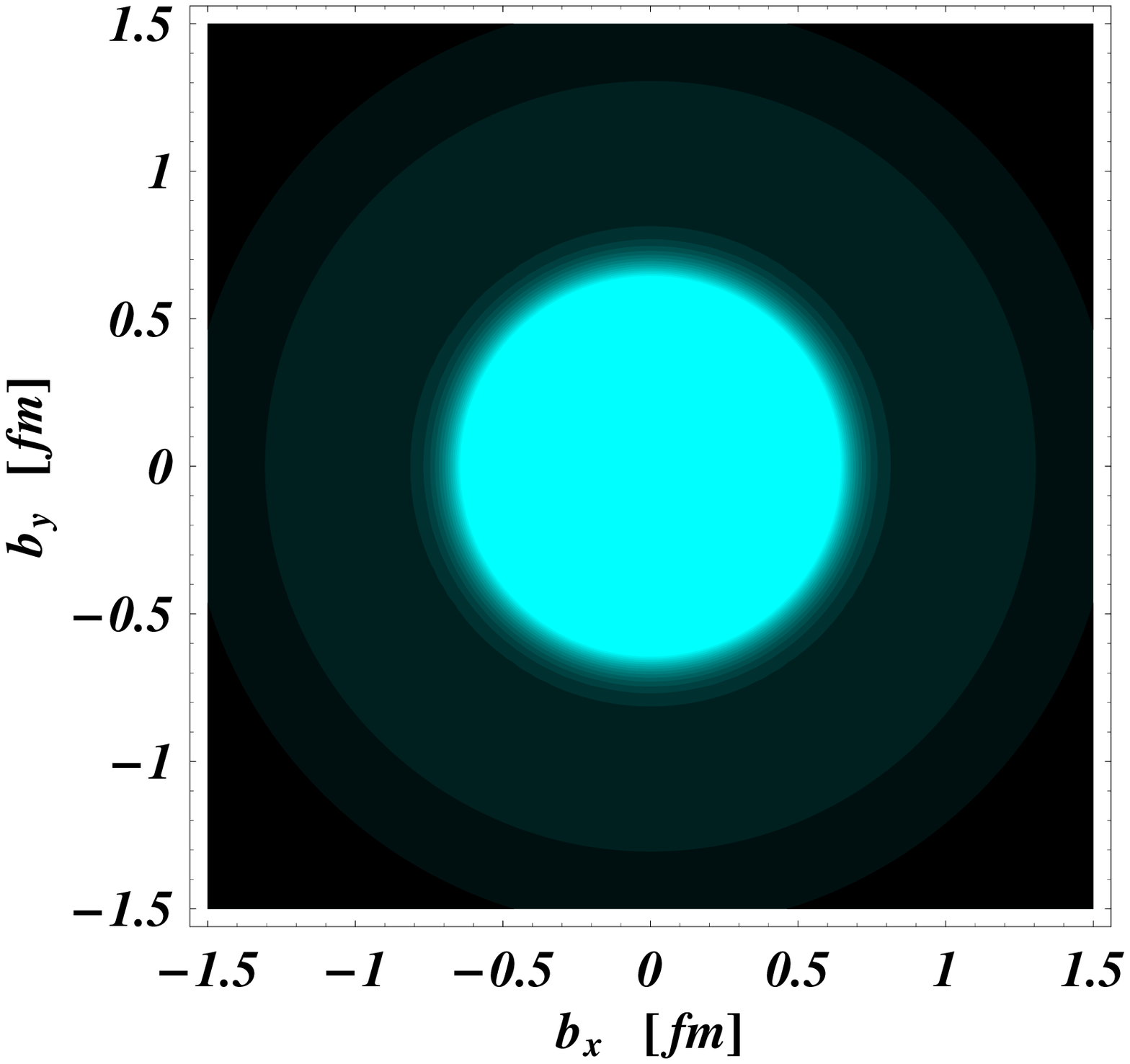}
\end{center}
\vspace{0.25cm}
\begin{center}
\includegraphics[width =10.cm]{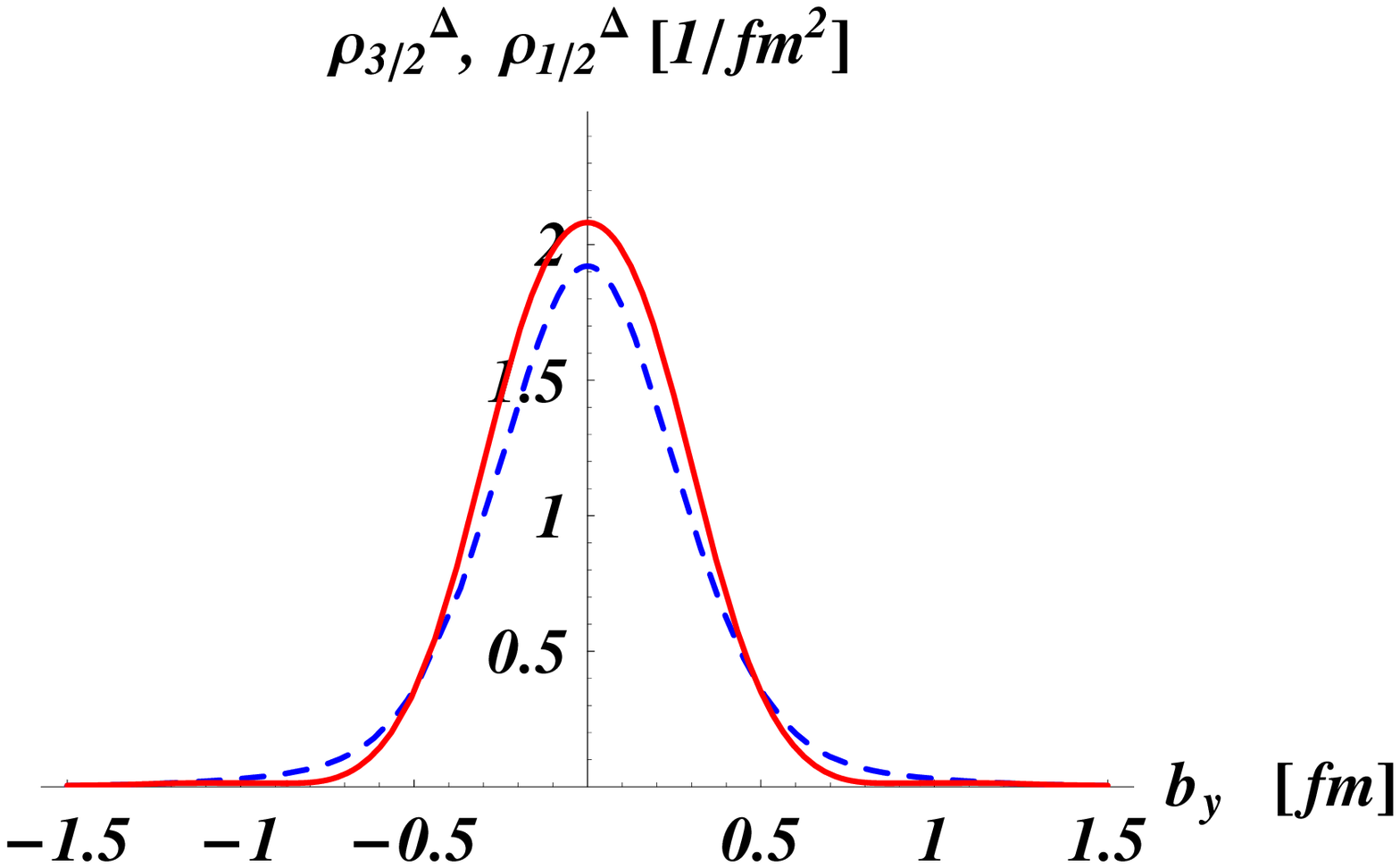}
\end{center}
\caption{Quark transverse charge densities in a {\it $\Delta^+(1232)$} of 
definite light-cone helicity. 
Upper left panel : $\rho^\Delta_{\frac{3}{2}}$. 
Upper right panel : $\rho^\Delta_{\frac{1}{2}}$. 
The light (dark) regions correspond to the
largest (smallest) values of the density. 
The lower panel compares the density along the $y$-axis 
for  $\rho^\Delta_{\frac{3}{2}}$ (dashed curve) and 
$\rho^\Delta_{\frac{1}{2}}$ (solid curve). 
For the $\Delta$ e.m. FFs, 
the quenched lattice QCD results are used 
(fit of Table~\ref{table:latticefit1}). }
\label{fig:deltahel}
\end{figure}

In Fig.~\ref{fig:deltahel}, we compare the $\Delta^+$ transverse densities 
in helicity states of $\lambda = 3/2$ and $\lambda = 1/2$, using 
Eq.~(\ref{eq:dens2}),  for the quenched 
lattice calculations. A comparison reveals that both are very similar, 
showing a positive core in the transverse charge density of about 0.7~fm. 
One notices that the
quark transverse density in a  $\lambda = 1/2$ state is slightly more 
concentrated than its $\lambda = 3/2$ counterpart. 

\begin{figure}[H]
\begin{center}
\includegraphics[width = 0.49 \linewidth]{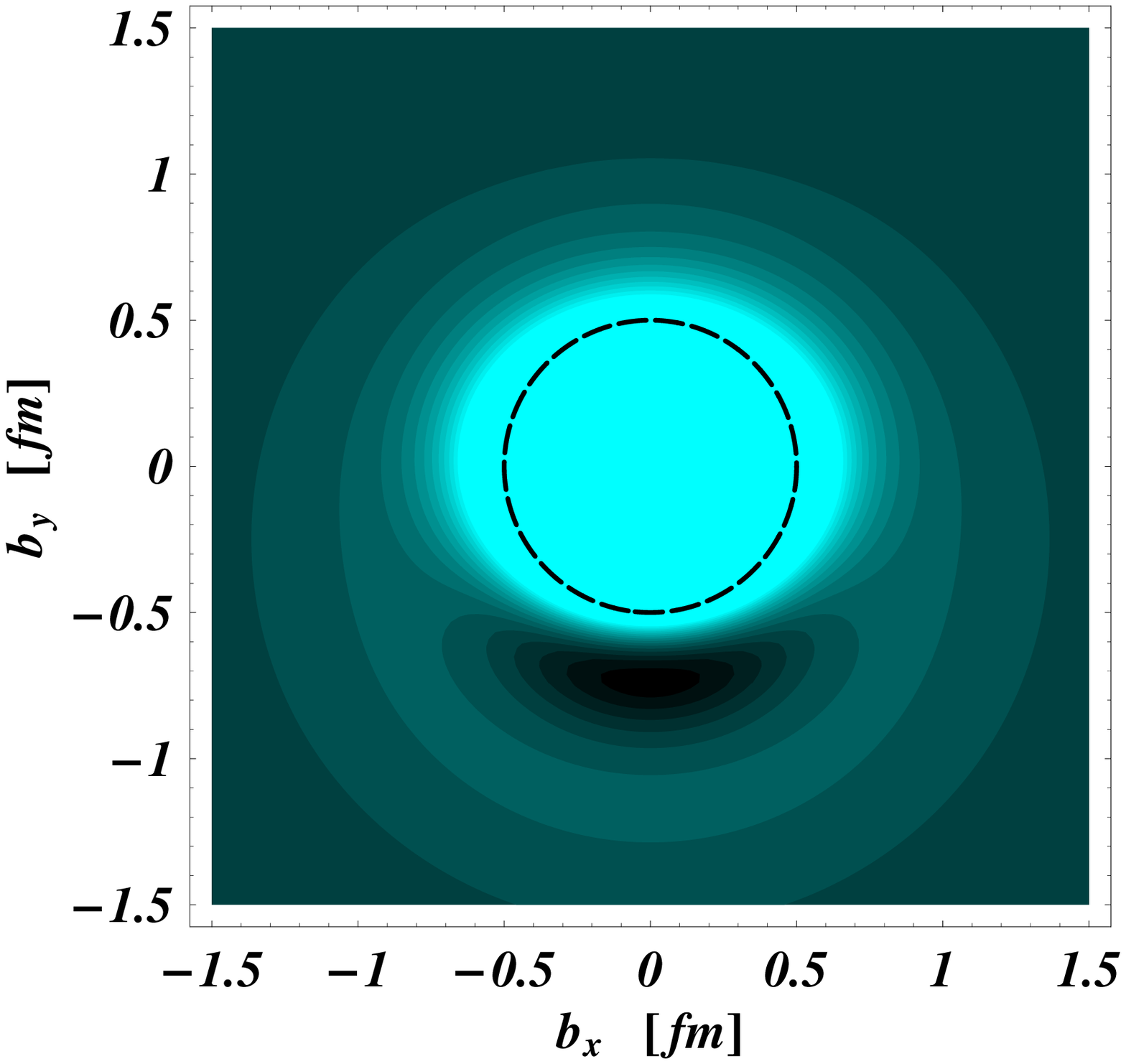}
\includegraphics[width = 0.49 \linewidth]{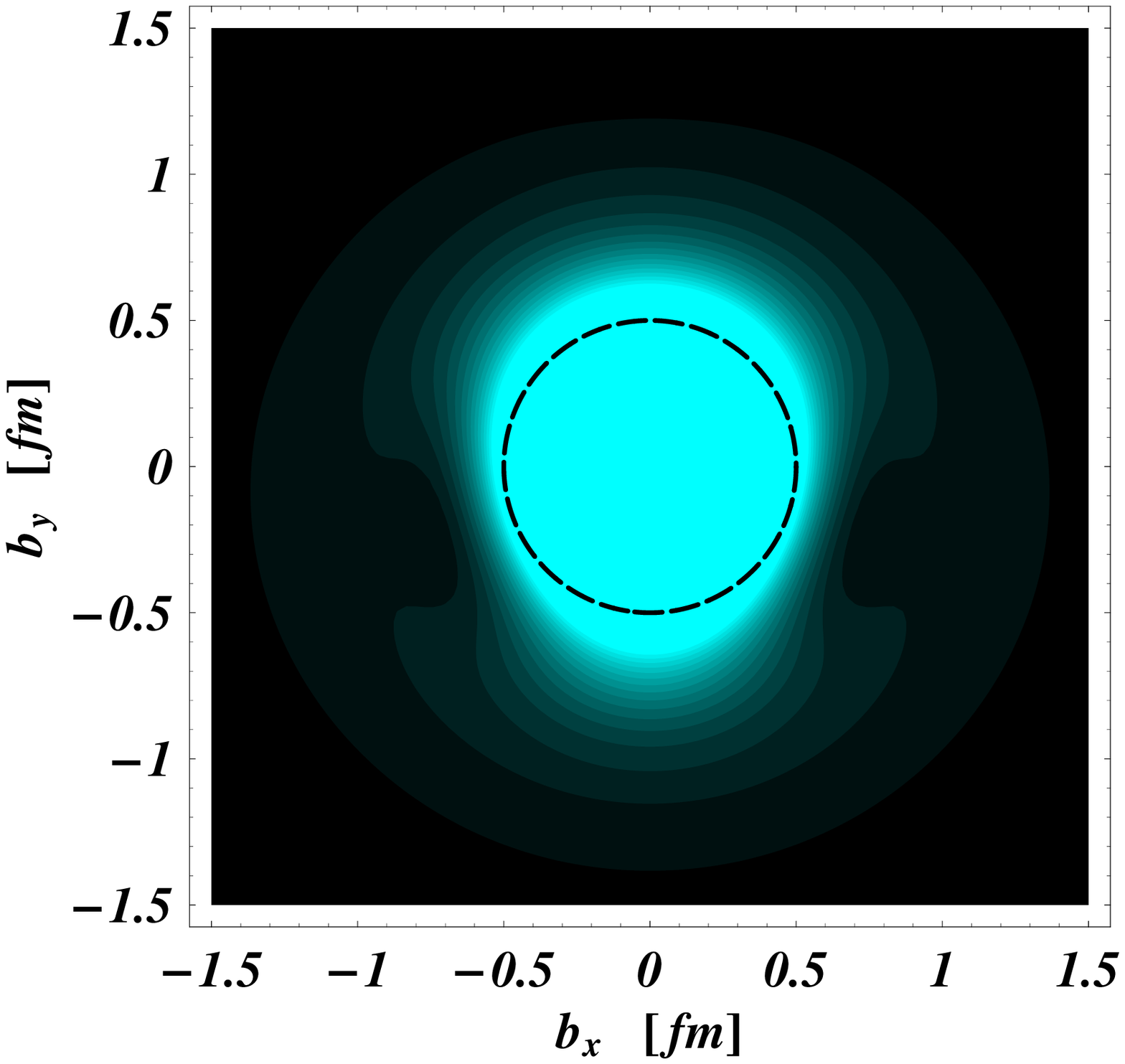}
\end{center}
\vspace{0.25cm}
\begin{center}
\includegraphics[width =10cm]{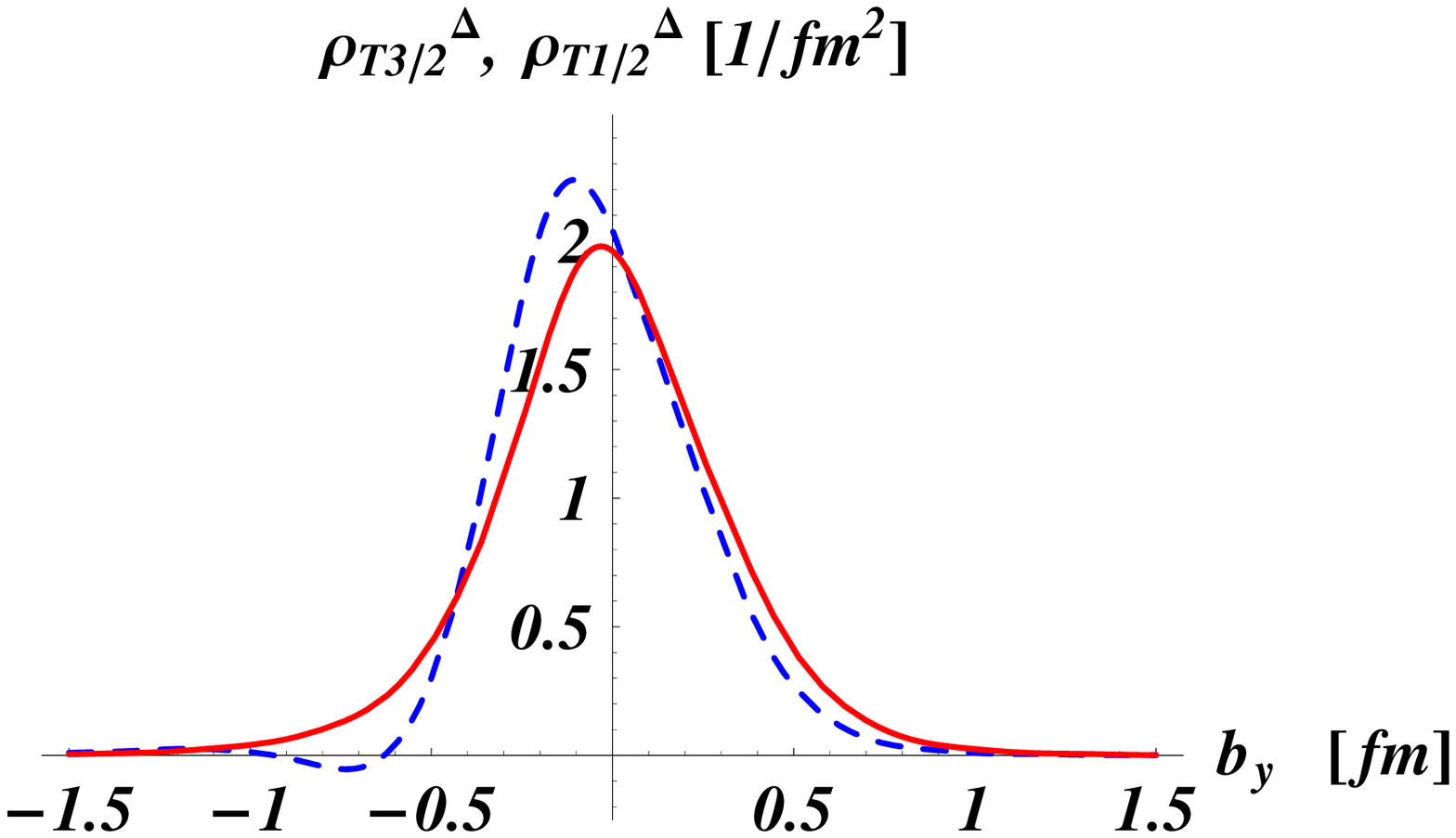}
\end{center}
\caption{Quark transverse charge densities in a 
{\it $\Delta^+(1232)$} which 
is polarized along the positive $x$-axis. 
Upper left panel : $\rho^\Delta_{T \, \frac{3}{2}}$. 
Upper right panel : $\rho^\Delta_{T \, \frac{1}{2}}$. 
The light (dark) regions correspond to the 
largest (smallest) values of the density. 
In order to see the deformation more clearly, a circle of radius 0.5~fm is drawn for 
comparison. 
The lower panel compares the density along the $y$-axis 
for  $\rho^\Delta_{T \, \frac{3}{2}}$ (dashed curve) and 
$\rho^\Delta_{T \, \frac{1}{2}}$ (solid curve). 
For the $\Delta$ e.m. FFs, 
the quenched lattice QCD results are used  
(fit of Table~\ref{table:latticefit1}). }
\label{fig:deltatrans}
\end{figure}

In Fig.~\ref{fig:deltatrans}, the transverse densities of 
Eqs.~(\ref{eq:dens4},\ref{eq:dens5}) are compared 
for a $\Delta^+$ that has a transverse spin. It is seen that the 
quark charge density in a 
$\Delta^+$ in a state of transverse spin projection 
$s_\perp = +3/2$ is elongated 
along the axis of the spin (prolate deformation)  
whereas in a state of transverse spin 
projection $s_\perp = +1/2$ it is elongated along the axis perpendicular to
the spin. 

\begin{figure}[H]
\begin{center}
\includegraphics[width = 0.49 \linewidth]{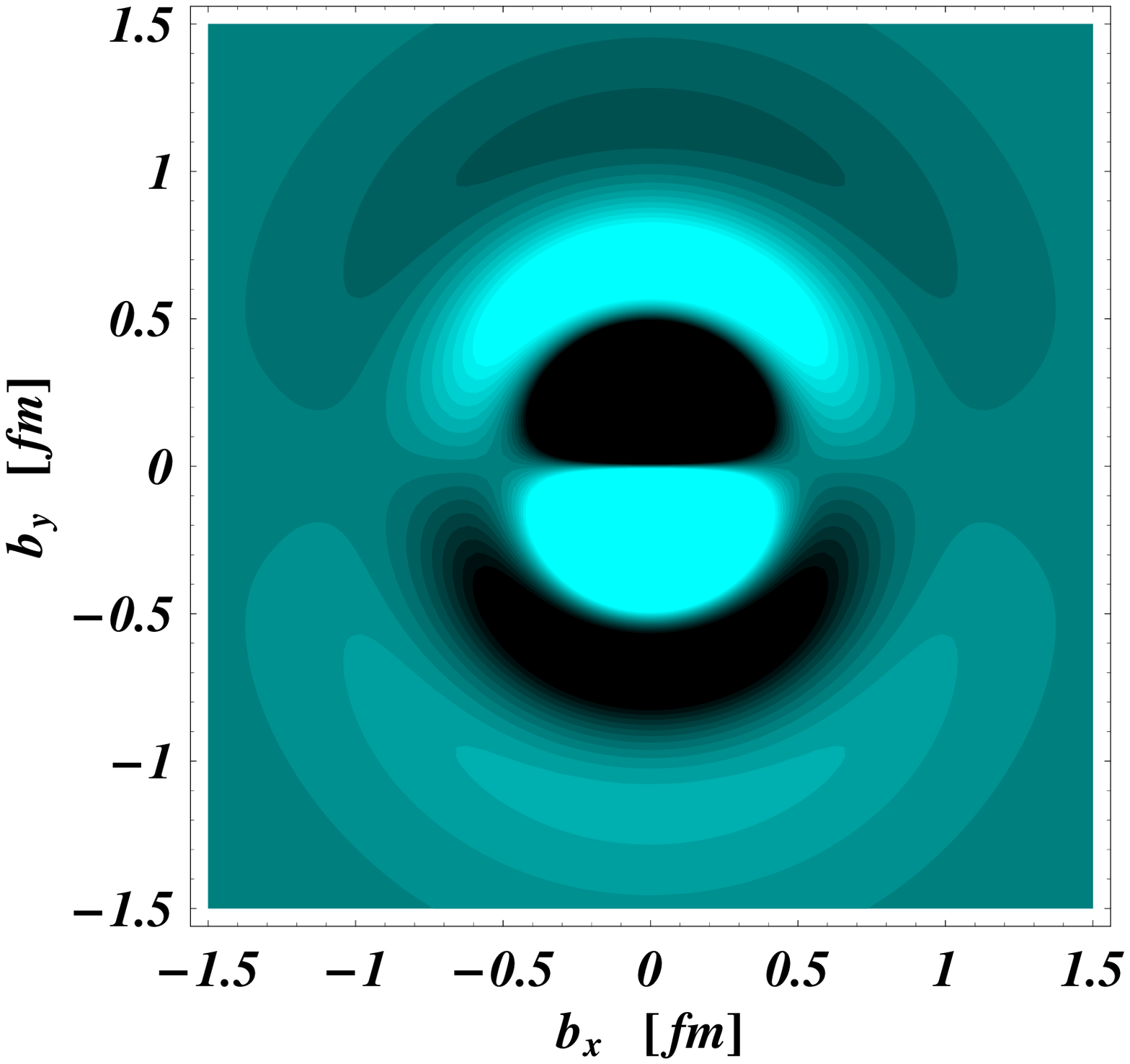}
\includegraphics[width = 0.49 \linewidth]{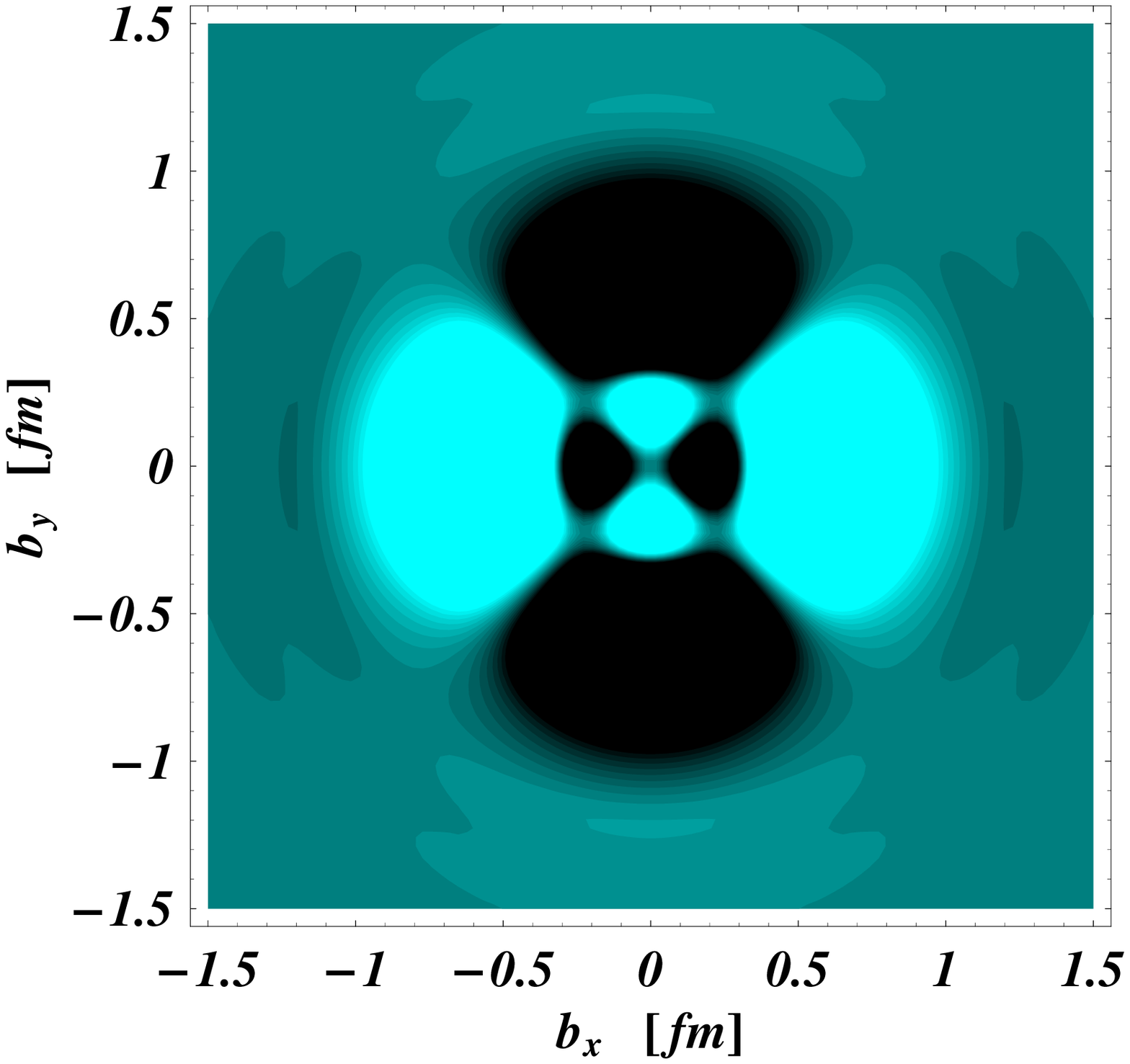}
\end{center}
\hspace{0.25cm}
\begin{center}
\includegraphics[width = 0.49 \linewidth]{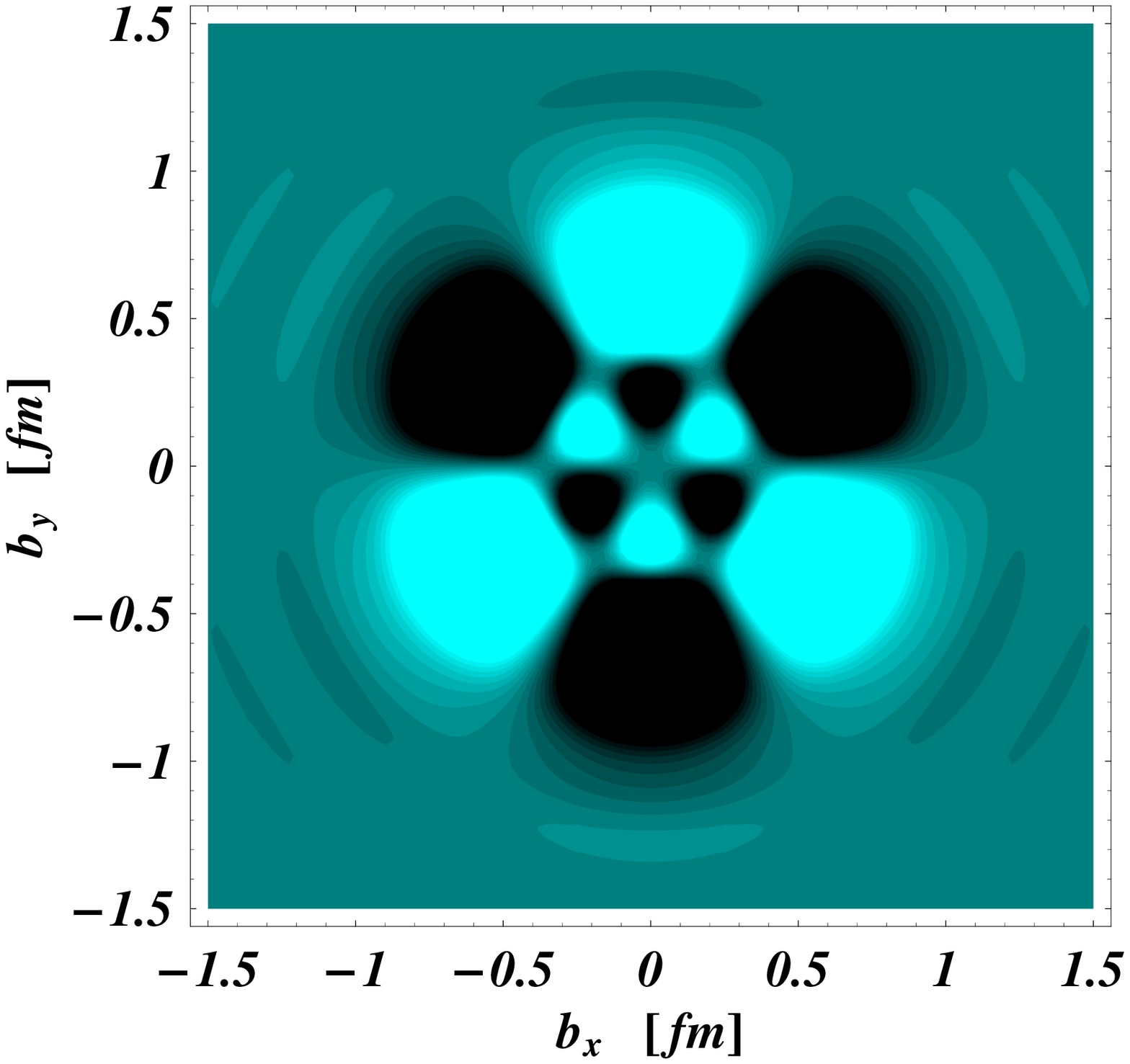}
\end{center}
\caption{Multipole field patterns in the 
quark transverse charge density $\rho^\Delta_{T \, \frac{3}{2}}$ 
in a {\it $\Delta^+(1232)$} that  
is polarized along the positive $x$-axis. 
Upper left panel : dipole field pattern; 
upper right panel : quadrupole field pattern; 
lower panel : octupole field pattern.  
For the $\Delta$ e.m. FFs,  
the quenched lattice QCD results are used 
(fit of Table~\ref{table:latticefit1}). }
\label{fig:deltatrans2}
\end{figure}

The corresponding dipole, quadrupole and octupole field patterns 
in the transverse quark charge density 
for a transversely polarized $\Delta$ are shown in 
Fig.~\ref{fig:deltatrans2}.

\begin{figure}[H]
\begin{center}
\includegraphics[width =10cm]{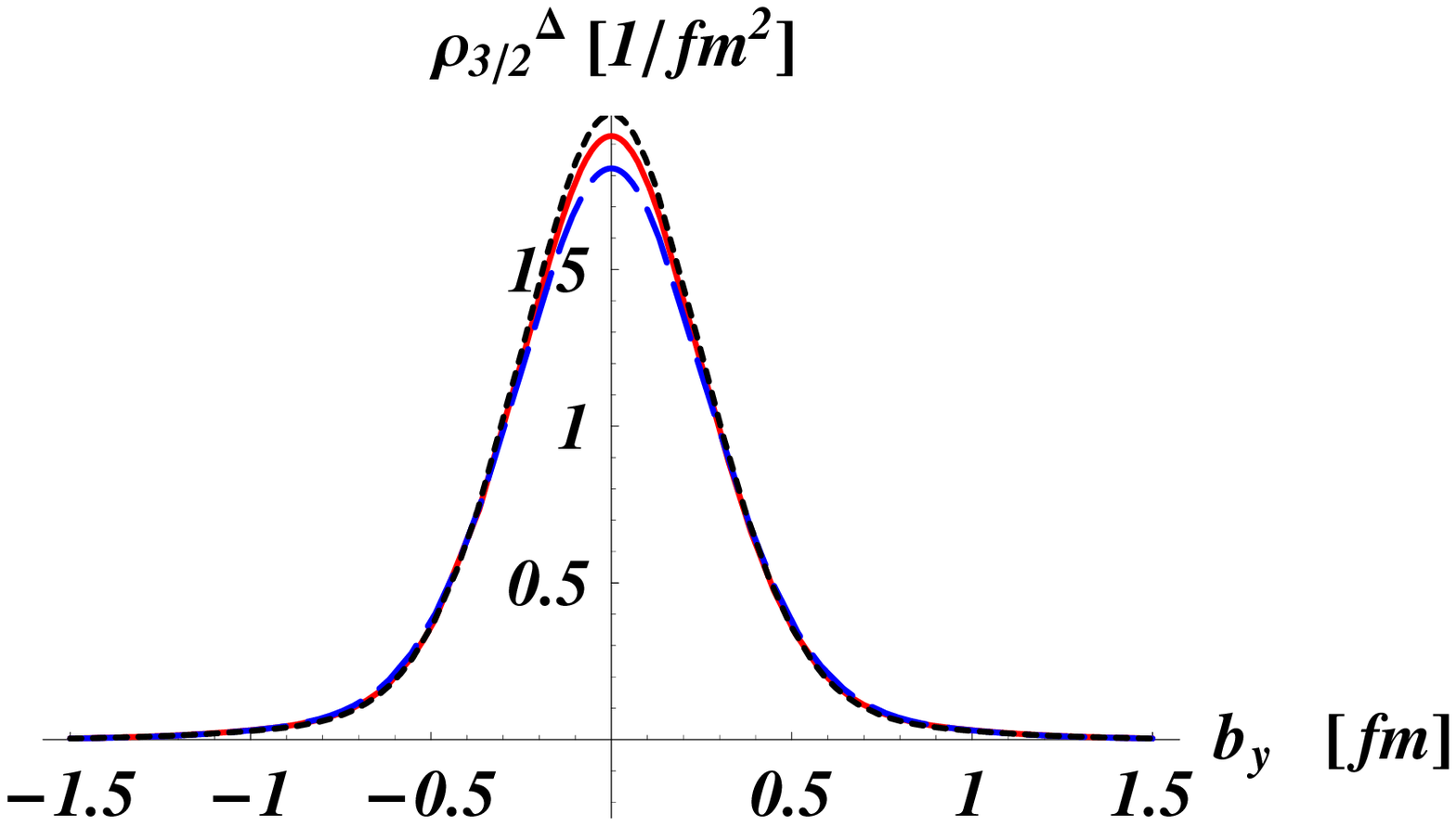}
\end{center}
\begin{center}
\includegraphics[width =10cm]{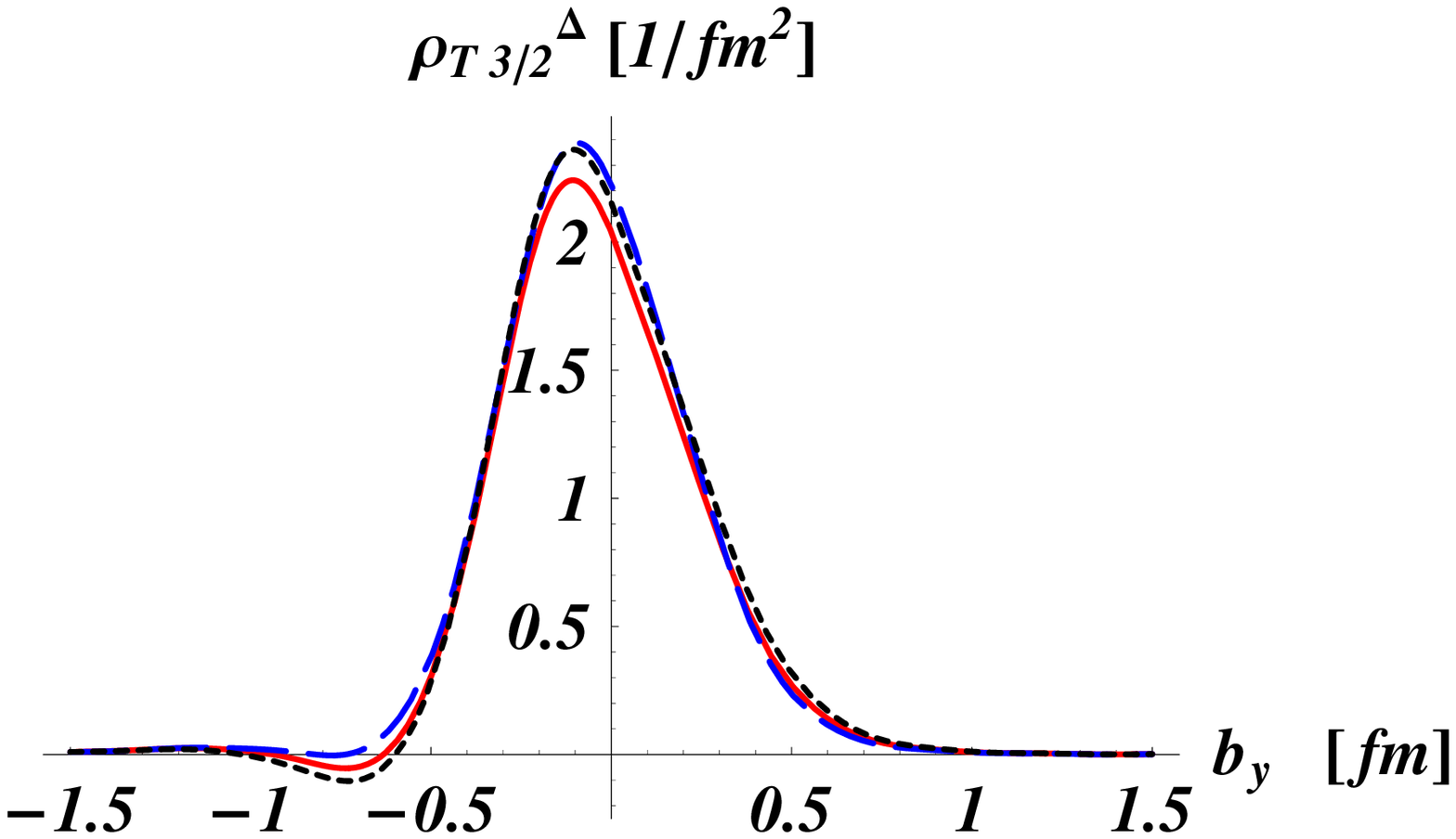}
\end{center}
\caption{
Comparison of the densities along the $y$-axis 
for  $\rho^\Delta_{\frac{3}{2}}$ (upper panel) and 
$\rho^\Delta_{T \, \frac{3}{2}}$ (lower panel), 
for three different QCD lattice calculations of the 
$\Delta^+(1232)$ e.m. FFs
according to the fit of Table~\ref{table:latticefit1}.  
Solid (red) curves~: quenched Wilson result; 
long dashed (blue) curves~: dynamical $N_f = 2$ Wilson result; 
dotted (black) curves~: hybrid lattice calculation. 
}
\label{fig:deltalatticecomp}
\end{figure}

To convey the consistency of all three types of simulations, 
we show the differences in the transverse densities between the 
quenched calculation and the two dynamical lattice calculations in 
Fig.~\ref{fig:deltalatticecomp}. One notices that the differences  are 
very small and arise primarily 
in the central densities for quenched and dynamical Wilson calculations. 

\section{Discussion and Conclusions}
\label{sec8}

In this paper we have studied the electromagnetic 
properties of the $\Delta(1232)$-resonance using lattice QCD.
Lattice results for the $\Delta$  electromagnetic
form factors have been presented for pion masses down to approximatively 350 MeV for  
three cases:
quenched QCD, two flavors of dynamical Wilson quarks, and three  
flavors of
quarks described by a mixed action combining domain wall valence quarks
and  staggered sea quarks.

We have also presented the formalism for understanding the spin-3/2 $\Delta$ form factors in terms of precisely defined transverse quark charge density distributions in the infinite momentum,  or
light-front, frame. For reference, we showed that a point-like transverse charge density, that is, a two dimensional delta function, yields the specific values $G_{E0}(0) =1$, $G_{M1}(0) = 3$, $G_{E2}(0) = -3 $, and $G_{M3}(0) = -1$ and that these values are the same as the `natural' values occurring for structureless gravitinos in extended supergravity. This strongly supports the hypothesis that these 
are universal values characterizing structureless spin-3/2 particles. The fact that our lattice calculations yield values significantly different from these `natural' values indicates the presence of interesting, nontrivial structure, and we have used the helicity amplitudes extracted from the electromagnetic form factors to construct the two-dimensional transverse quark charge densities for longitudinally and transversely polarized spins.  A salient feature is that for the state with transverse spin projection 3/2, the transverse quark charge distribution is prolately deformed along the axis of the spin.

Since there is also a very different language for describing the internal deformation of the $\Delta$ in terms of a three-dimensional intrinsic state density, it is useful to briefly summarize the status of model studies using that description and discuss the differences between the two languages. 
In nuclear physics, for which the nucleus is sufficiently massive that for momentum transfers characteristic of the spatial size of the nucleus, electromagnetic form factors are well described by the Fourier transform of a three-dimensional charge density, there is a fruitful history of determining the deformations of intrinsic states. The basic idea~\cite{bm} is to treat collective rotations of a well-deformed nucleus in terms of a deformed intrinsic state that is axially symmetric in a body fixed-frame in which the projection of the total angular momentum $J$ along the body-fixed symmetry axis is $K$. Following Ref.~\cite{bm}, the relation between the intrinsic quadrupole moment, $Q_0$,  and the spectroscopic quadrupole moment, $Q$, measured experimentally and corresponding to the Fourier transform of the density in the lab frame, is
\begin{equation}
Q = \frac{3K^2 -J(J+1)}{(J+1)(2J+3)} Q_0 \, .
\label{intrinsic}
\end{equation}
In nuclear physics, this approximation is not only successful phenomenologically, but also {\it ab initio} nuclear mean-field theory calculations using many-body theory that accurately predict electromagnetic form factors of spherical nuclei also predict the form factors arising from deformed intrinsic states in 
rare-earth and actinide nuclei~\cite{Negele:1977zz}.

It is natural to ask whether this intrinsic state approximation is also applicable in studying the nucleon. One obvious example is the Skyrme model~\cite{Skyrme:1961vq,Skyrme:1962vh}, in which a single intrinsic soliton state  generates an infinite band of collective spin-isopin rotations, with the spin-1/2 isospin-1/2 nucleon being the lowest state and the spin-3/2 isospin-3/2 $\Delta$ being the first excited state. It is clear, however,  that the Skryme model cannot be a quantitative model of the nucleon and $\Delta$, since the infinite tower of predicted additional states are not observed in Nature. Buchmann and Henley~\cite{Buchmann:2001gj} reviewed a variety of models of the $\Delta$, and concluded that for a quark model, a collective model, and a pion cloud model, the intrinsic state was oblate, corresponding to a negative value for $Q_0$.  Studying a spheriodal bag model, Viollier, Chin, and Kerman~\cite{Viollier:1983wy} concluded that the $K = 3/2$ intrinsic state (the same as considered by Buchmann and Henley) is oblate and that the $K=1/2$ intrinsic state is prolate (as in the Skyrme Model).  
We note that for a $J=3/2$  $\Delta$,  Eq.~(\ref{intrinsic}) specifies that for a $K=3/2$ intrinsic state $Q$ and $Q_0$ have the same sign whereas for a $K=1/2$ state they have opposite signs. Thus, both a prolate $K=1/2$ intrinsic state and oblate $K=3/2$ state yield a negative spectroscopic quadrupole moment, and indeed, all the results in Refs.~\cite{Buchmann:2001gj,Viollier:1983wy} are consistent with a negative value for $Q$ and, thus by Eq.~(\ref{eq:ee8c}),  with a negative value for $G_{E2}$ as we have calculated in Table~\ref{table:latticefit1}. A recent calculation of the $\Delta^+$ form factors in the chiral quark soliton model~\cite{Ledwig:2008es} found a value $G_{E2}(0) = -2.145$, consistent with the value of the hybrid lattice result shown in Table~\ref{table:latticefit1}. 

Unfortunately, it is not possible to go further and establish a correspondence between our result of a prolate deformation of the 2-dimensional transverse spin density along the axis of the spin-3/2 state and the model statement of an oblate intrinsic 3-dimensional density for a $K=3/2$ intrinsic state.  As we have emphasized, even a structureless $\Delta$ would have $G_{E2} = -3 $, so a negative  $G_{E2} $ less negative than $-3$ would indicate prolate deformation in our case while still being consistent with oblate deformation for an intrinsic state.  It is possible that study of a simple relativistic model that could be boosted into the infinite momentum frame could go somewhat further in bridging the gap between these two different descriptions. 
 
 \section*{Acknowledgments} 
It is a pleasure to acknowledge helpful discussions with Ernest Henley and Jerry Miller. 
This work is
supported in part by the Cyprus Research Promotion Foundation (RPF) 
under contract $\Pi$ENEK/ENI$\Sigma$X/0505-39, 
the  EU Integrated Infrastructure Initiative Hadron Physics (I3HP) 
under contract RII3-CT-2004-506078 and 
the U.S. Department of Energy (D.O.E.) Office of Nuclear Physics 
under contracts DE-FG02-94ER40818 and DE-FG02-04ER41302.
This research used computational resources provided by RFP 
under contract EPYAN/0506/08, 
the National Energy Research Scientific Computing  Center 
supported by the Office of Science of the U.S. Department of Energy 
under Contract DE-AC03-76SF00098
and 
the MIT Blue Gene computer under grant DE-FG02-05ER25681.
Dynamical staggered quark configurations and forward domain wall quark 
propagators were provided by the MILC and LHPC collaborations respectively.

\appendix
\section{Lattice results for the Delta electromagnetic form factors}

In Tables~\ref{quenched-results}, \ref{wilson-results}, and
\ref{hybrid-results}, we list the lattice results for the Delta e.m. FFs in
the quenched, dynamical Wilson, and hybrid calculations respectively.  

\begin{table}[H]
\small
\begin{tabular}{l l l l l}
\hline\hline 
& $Q^2$ (GeV$^2$) & $G_{E0}$ & $G_{M1}$ & $G_{E2}$ \\
\hline
&0.173093(69) & 0.7750(27) & 2.340(47) &   \\
&0.33976(26)  & 0.6228(41) & 1.913(40) & -0.37(13) \\
&0.50068(54)  & 0.5136(54) & 1.600(36) & -0.250(97) \\
&0.65640(89)  & 0.4368(68) & 1.361(32) &   \\
&0.8074(13)   & 0.3728(69) & 1.182(29) & -0.216(72) \\
$m_\pi= 0.563(4)$ GeV &0.9541(18)   & 0.3205(75) & 1.027(29) & -0.170(60) \\
&1.2358(28)   & 0.2527(85) & 0.809(31) & -0.177(57) \\
&1.3715(34)   & 0.2174(87) & 0.701(31) & -0.159(50) \\
&1.5040(40)   & 0.194(10)  & 0.628(33) & -0.084(63) \\
&1.6335(46)   & 0.1698(97) & 0.549(31) & -0.080(50) \\
&1.7603(53)   & 0.142(11)  & 0.460(38) & -0.136(57) \\
&1.8845(59)   & 0.140(12)  & 0.439(35) & -0.100(55) \\
&2.0063(66)   & 0.117(10)  & 0.372(31) & -0.085(44) \\
&2.2430(79)   & 0.084(16)  & 0.284(47) &   \\
\hline
&0.172859(82) & 0.7646(32) & 2.324(58) &   \\
&0.33890(30)  & 0.6076(46) & 1.889(48) & -0.41(16) \\
&0.49887(64)  & 0.4965(61) & 1.575(43) & -0.27(12) \\
&0.6534(11)   & 0.4212(77) & 1.329(37) &   \\
&0.8030(15)   & 0.3574(75) & 1.155(34) & -0.209(85) \\
$m_\pi= 0.490(4)$ GeV &0.9481(21)   & 0.3055(83) & 1.004(33) & -0.160(70) \\
&1.2263(33)   & 0.2428(90) & 0.797(35) & -0.196(68) \\
&1.3600(40)   & 0.2060(93) & 0.682(35) & -0.167(61) \\
&1.4906(47)   & 0.182(11)  & 0.608(37) & -0.054(76) \\
&1.6181(54)   & 0.159(10)  & 0.531(34) & -0.062(59) \\
&1.7427(61)   & 0.131(12)  & 0.440(41) & -0.151(72) \\
&1.8648(68)   & 0.133(12)  & 0.432(39) & -0.118(65) \\
&1.9844(76)   & 0.110(10)  & 0.358(33) & -0.091(50) \\
&2.2167(91)   & 0.073(16)  & 0.259(47) &   \\ 
\hline
&0.172616(98) & 0.7546(42) & 2.320(78) &   \\
&0.33800(36)  & 0.5933(55) & 1.876(62) & -0.50(22) \\
&0.49700(75)  & 0.4797(72) & 1.558(54) & -0.32(15) \\
&0.6503(12)   & 0.4061(92) & 1.295(48) &   \\
&0.7985(18)   & 0.3422(84) & 1.131(41) & -0.20(12) \\
$m_\pi= 0.311(4)$ GeV &0.9420(24)   & 0.2908(95) & 0.987(41) & -0.170(87) \\
&1.2167(39)   & 0.236(10)  & 0.797(41) & -0.244(90) \\
&1.3485(46)   & 0.196(10)  & 0.665(41) & -0.199(83) \\
&1.4770(54)   & 0.172(12)  & 0.590(43) & -0.03(10) \\
&1.6025(62)   & 0.150(11)  & 0.518(39) & -0.041(79) \\
&1.7250(70)   & 0.121(13)  & 0.417(48) & -0.19(11) \\
&1.8450(79)   & 0.127(13)  & 0.436(46) & -0.193(88) \\
&1.9624(87)   & 0.103(11)  & 0.349(37) & -0.117(62) \\
&2.190(10)    & 0.066(17)  & 0.231(52) &   \\ 
\hline\hline
\end{tabular}
\caption{Delta form factors from quenched Wilson fermions.}
\label{quenched-results}
\end{table}

\begin{table}[H]
\small
\begin{tabular}{l l l l l}
\hline\hline 
& $Q^2$ (GeV$^2$) & $G_{E0}$ & $G_{M1}$ & $G_{E2}$ \\
\hline
&0.43260(36) & 0.6248(40) & 1.900(50) &   \\
&0.8364(13)  & 0.4352(65) & 1.348(44) & -0.33(15) \\
&1.2164(25)  & 0.337(11)  & 0.998(49) & -0.17(12) \\
&1.5765(40)  & 0.282(19)  & 0.761(56) &   \\
$m_\pi= 0.691(8)$ GeV &1.9195(56)  & 0.235(22)  & 0.638(58) & -0.12(10) \\
&2.2475(74)  & 0.189(21)  & 0.501(61) & -0.095(79) \\
&2.866(11)   & 0.125(32)  & 0.312(84) & -0.012(96) \\
&3.159(13)   & 0.103(24)  & 0.225(56) & -0.043(75) \\
&3.442(15)   & 0.069(19)  & 0.136(43) & -0.155(96) \\
&3.717(17)   & 0.054(27)  & 0.129(58) & -0.04(11) \\
&3.984(19)   & 0.081(47)  & 0.106(70) & -0.12(18) \\
&4.244(21)   & 0.034(13)  & 0.087(32) & -0.008(58) \\
&4.497(23)   & 0.040(33)  & 0.078(72) & -0.010(93) \\
&4.985(27)   & 0.019(34)  & -0.004(49)&   \\
\hline
&0.42932(57) & 0.5748(85) & 1.743(76) &   \\
&0.8250(20)  & 0.3785(82) & 1.199(55) & -0.49(21) \\
&1.1939(39)  & 0.288(13)  & 0.745(60) & -0.41(19) \\
&1.5409(61)  & 0.241(30)  & 0.546(78) &   \\
$m_\pi= 0.509(8)$ GeV &1.8694(85)  & 0.181(16)  & 0.433(55) & -0.15(12) \\
&2.182(11)   & 0.194(51)  & 0.36(11)  & -0.04(13) \\
&2.768(16)   & 0.093(33)  & 0.181(75) & -0.22(14) \\
&3.044(19)   & 0.037(18)  & 0.043(37) & 0.004(91) \\
&3.311(22)   & 0.019(15)  & 0.039(26) & -0.021(84) \\
&3.569(25)   & 0.021(39)  & 0.032(62) & -0.06(14) \\
&3.819(27)   & 0.045(41)  & -0.009(59)& 0.05(17) \\
&4.062(30)   & 0.013(16)  & 0.020(29) & -0.018(74) \\
&4.298(33)   & 0.003(19)  & -0.001(27)& 0.099(82) \\
&4.753(38)   & 0.015(43)  & -0.00(11) &   \\
\hline
&0.42485(79) & 0.566(15) & 1.530(92) &   \\
&0.8099(26)  & 0.368(16) & 1.048(62) & -0.53(27) \\
&1.1647(50)  & 0.276(32) & 0.891(97) & -0.41(24) \\
$m_\pi= 0.384(8)$ GeV&1.4953(77)  & 0.195(48) & 0.57(11)  &   \\
 &1.806(11)   & 0.141(80) & 0.354(39) & -0.30(15) \\
&2.100(14)   & 0.11(12)  & 0.337(61) & -0.27(15) \\
&2.648(20)   & 0.11(25)  & 0.062(99) & -0.42(49) \\
&2.905(23)   & 0.05(31)  & 0.095(37) & 0.04(12) \\
&3.152(26)   & 0.04(39)  & 0.084(53) & -0.48(24) \\
&3.391(29)   & 0.03(47)  & 0.047(95) & -0.19(39) \\
&3.622(32)   & 0.04(57)  & -0.040(94)& 0.17(23) \\
&3.846(35)   & 0.03(68)  & 0.001(54) & -0.05(20) \\
&4.063(38)   & 0.04(78)  & 0.014(35) & -0.06(10) \\
&4.481(44)   & -0.0(1.0) & 0.011(69) &   \\
\hline\hline
\end{tabular}
\caption{Delta form factors from dynamical Wilson fermions.}
\label{wilson-results}
\tablab{wilson-results}
\end{table}

\begin{table}[H]
\small
\begin{tabular}{l l l l l}
\hline\hline 
& $Q^2$ (GeV$^2$) & $G_{E0}$ & $G_{M1}$ & $G_{E2}$ \\
\hline
&0.124094(51) & 0.8223(95) & 2.83(21)  &   \\
&0.24512(19)  & 0.686(13)  & 2.26(17)  & -1.47(88) \\
&0.36330(42)  & 0.584(17)  & 1.99(17)  & -0.84(72) \\
&0.47881(71)  & 0.499(22)  & 1.89(17)  &   \\
&0.5918(11)   & 0.432(21)  & 1.53(14)  & -0.76(48) \\
&0.7025(15)   & 0.386(23)  & 1.38(14)  & -0.56(46) \\
&0.9175(24)   & 0.310(29)  & 1.15(14)  & -0.20(44) \\
&1.0220(29)   & 0.287(35)  & 1.04(15)  & -0.47(43) \\
$m_\pi= 0.353(2)$ GeV &1.1246(35)   & 0.239(31)  & 0.83(13)  & -0.05(47) \\
&1.2255(41)   & 0.223(34)  & 0.78(14)  & -0.44(47) \\
&1.3247(47)   & 0.261(88)  & 0.90(31)  & -0.72(65) \\
&1.4223(53)   & 0.187(35)  & 0.67(12)  & -0.38(43) \\
&1.5184(59)   & 0.176(45)  & 0.65(16)  & -0.81(48) \\
&1.7064(73)   & 0.094(42)  & 0.27(15)  &   \\
&1.7983(80)   & 0.125(45)  & 0.41(13)  & -0.64(48) \\
&1.8890(87)   & 0.095(34)  & 0.327(85) & -0.40(39) \\
&1.9785(94)   & 0.101(40)  & 0.30(13)  & -0.84(64) \\
&2.067(10)    & 0.073(35)  & 0.276(87) & -0.39(44) \\
&2.154(11)    & 0.073(37)  & 0.191(84) & -0.51(43) \\
&2.240(12)    & 0.124(89)  & 0.23(27)  & -1.3(1.4) \\
&2.409(13)    & 0.079(55)  & 0.16(16)  & -0.45(71) \\
\hline\hline
\end{tabular}
\caption{Delta form factors from the hybrid approach.}
\label{hybrid-results}
\tablab{hybrid-results}
\end{table}

\end{document}